\documentclass[aps,onecolumn,superscriptaddress,long,notitlepage,letterpaper,balancelastpage,nofootinbib,prd,floatfix]{revtex4-2}
\pdfoutput=1

\usepackage{amsmath,mathtools,amssymb,amsthm,amsxtra,overpic,bbm,epsfig,subfigure,url,bm}
\usepackage{hyperref}
\usepackage{mathrsfs}
\usepackage[dvipsnames]{color,xcolor}
\usepackage{comment}
\usepackage{float}
\usepackage{enumitem}
\usepackage{slashed}
\usepackage{bbold}

\setlist{nolistsep}

\definecolor{nicered}{rgb}{0.5,0.,0.}
\definecolor{nicegreen}{rgb}{0.,0.5,0.}
\definecolor{niceblue}{rgb}{0.,0.,0.5}
\hypersetup{
colorlinks=true,
linkcolor=black,
filecolor=nicegreen,      
urlcolor=niceblue,
citecolor=nicered,
}

\begin{document}
\title{Cosmology-friendly time-varying neutrino masses via the sterile neutrino portal}
\author{Guo-yuan Huang}
\email{guoyuan.huang@mpi-hd.mpg.de}
\affiliation{Max-Planck-Institut f{\"u}r Kernphysik, Saupfercheckweg 1, 69117 Heidelberg, Germany}

\author{Manfred Lindner}
\email{manfred.lindner@mpi-hd.mpg.de} 
\affiliation{Max-Planck-Institut f{\"u}r Kernphysik, Saupfercheckweg 1, 69117 Heidelberg, Germany}

\author{Pablo Mart\'{i}nez-Mirav\'e}
\email{pamarmi@ific.uv.es}
\affiliation{Departament de F\'isica  Te\'orica,  Universitat  de  Val\`{e}ncia, and Instituto de F\'{i}sica Corpuscular, CSIC-Universitat de Val\`{e}ncia, 46980 Paterna, Spain}

\author{Manibrata Sen}
\email{manibrata@mpi-hd.mpg.de}
\affiliation{Max-Planck-Institut f{\"u}r Kernphysik, Saupfercheckweg 1, 69117 Heidelberg, Germany}

%%%%%%%%%%%%%%%%%%%%%%%%%%
\begin{abstract}
	\noindent
	We investigate a consistent scenario of time-varying neutrino masses, and discuss its impact on cosmology, beta decay, and neutrino oscillation experiments.
	Such time-varying masses are assumed to be generated by the coupling between a sterile neutrino and an ultralight scalar field, which in turn affects the light neutrinos by mixing. Besides, the scalar could act as an ultralight dark matter candidate.
	We demonstrate how various cosmological bounds, such as those coming from Big Bang nucleosynthesis, the cosmic microwave background, as well as large scale structures, can be evaded in this model. 
	This scenario can be further constrained using multiple terrestrial experiments.
	In particular, for beta-decay experiments like KATRIN, non-trivial distortions to the electron spectrum can be induced, even when time-variation is fast and it gets averaged. 
	Furthermore, the presence of time-varying masses of sterile neutrinos will alter the interpretation of light sterile neutrino parameter space in the context of the reactor and gallium anomalies.
	In addition, we also study the impact of such time-varying neutrino masses on results from the BEST collaboration, which have recently strengthened the gallium anomaly. If confirmed, we find that the time-varying neutrino mass hypothesis could give a better fit to the recent BEST data.
\end{abstract}

\maketitle
\vspace{-15pt}

\section{Introduction}
\noindent
%%%%%%%%%%
%%%%%%%%
Ever since its first detection in 1956~\cite{Cowan:1956rrn,Reines:1953pu}, the elusive neutrinos remain to be the least known fermion in the Standard Model (SM)~\cite{Athar:2021xsd,ParticleDataGroup:2020ssz}.
Though the early beta-decay data suggest neutrino mass to be either vanishing or extremely small compared to the electron mass~\cite{Fermi:1934hr, Fermi:1934sk,Perrin:1933,Robertson:1991vn,Holzschuh:1992np,Kawakami:1991th,Sun:1993,Stoeffl:1995wm}, the discovery of neutrino oscillation phenomenon has firmly established the fact that neutrinos are massive.
To weigh those massive neutrinos in a model-independent manner is the major task of modern beta-decay experiments~\cite{Kraus:2004zw,Troitsk:2011cvm,KATRIN:2019yun,KATRIN:2021fgc,KATRIN:2021uub,KATRIN:2022ayy,Project8:2022wqh}.

It remains a mystery as to why the absolute scale of neutrino masses is more than six orders of magnitude smaller than its charged lepton partner. This is often ascribed to some new physics at very high energy scales in the spirit of the seesaw mechanism~\cite{Ma:1998dn,Cai:2017jrq,Cai:2017mow,Fritzsch:1975sr,Cheng:1975gk,Fritzsch:1975rz,Minkowski:1977sc,Yanagida:1980xy,GellMann:1980vs,Mohapatra:1979ia,Mohapatra:1980yp,Lazarides:1980nt,Konetschny:1977bn,Magg:1980ut,Schechter:1980gr,Cheng:1980qt,Mohapatra:1980yp,Lazarides:1980nt,Foot:1988aq}, whose experimental test is extremely challenging.
Alternatively, an interesting possibility is to attribute the origin of neutrino masses to the dark sectors, for example, dark energy (DE)~\cite{Frieman:1995pm,Fardon:2003eh} and dark matter (DM)~\cite{Davoudiasl:2018hjw}. As the persistent direct detection searches of weakly interacting massive particles (WIMPs) come out with null signals so far, there has been increasing attention to other DM candidates, such as the ultralight DM~\cite{Ferreira:2020fam,Urena-Lopez:2019kud}.
The ultralight DM candidate, due to its tiny mass $m<\mathcal{O}({\rm eV})$, acts as a delocalized \emph{classical-number} field.
Such a scenario could be responsible for addressing generic studies of varying physical constants.
One particular class of ultralight DM, the fuzzy dark matter with a mass $m^{}_{\phi}\gtrsim 10^{-22}~{\rm eV}$, can also help to alleviate the small-scale structure issues existing between cosmological observations and simulations \cite{Moore:1994yx,Flores:1994gz,Navarro:1996gj,Klypin:1999uc}.

Such an ultralight scalar field, coupled to neutrinos, can give rise to intriguing phenomenological consequences.
Neutrinos coupled to the scalar will naturally get a contribution to their masses, in analogy to the Higgs mechanism. As the classical DM field oscillates, the generated mass term for the neutrinos becomes time-varying.
There have been various attempts of studying the phenomenology of a coupling between neutrinos and the ultralight field in the literature~\cite{Berlin:2016woy,Brdar:2017kbt,Krnjaic:2017zlz,Zhao:2017wmo,Liao:2018byh,Capozzi:2018bps,Reynoso:2016hjr,Huang:2018cwo,Pandey:2018wvh,Farzan:2018pnk,Farzan:2019yvo,Choi:2019ixb,Baek:2019wdn,Ge:2019tdi,Choi:2019zxy,Choi:2020ydp,Dev:2020kgz,Baek:2020ovw,Losada:2021bxx,Smirnov:2021zgn,Alonso-Alvarez:2021pgy,Huang:2021zzz,Huang:2021kam,Chun:2021ief,Reynoso:2022vrn,Dev:2022bae}.
A simple realization of this possibility is to couple a classical scalar field ($\Phi$) to neutrinos via the operator $\overline{\nu}\nu \Phi$, which  generates a time-varying neutrino mass term.
In a UV-complete model, this can be achieved by directly coupling $\Phi$ to the lepton doublet $\ell^{}_{\rm L} \equiv (\nu, e)_{\rm L}$, as in the type-II seesaw~\cite{Konetschny:1977bn,Magg:1980ut,Schechter:1980gr,Cheng:1980qt}.
However, stringent constraints arise because of the accurate determination of electron properties.

A more feasible way to couple light neutrinos to the scalar field is by mixing with gauge singlets, such as a right-handed sterile neutrino $N$.
It is also natural to have a considerable Yukawa coupling between the sterile component and the scalar field.
Motivated by this, our working Lagrangian consistent with gauge symmetries is  described as,
%%%%%%%%%%%%%%%%%%%
\begin{eqnarray}
-\mathcal{L}& \supset &   y^{}_{\rm D} \overline{\ell^{}_{\rm L}} \widetilde{h} N +  \frac{1}{2}(m^{}_{N} + g \Phi) \overline{N^{\rm c}_{}}N  + \frac{1}{2} \kappa \overline{\ell^{}_{\rm L}} \widetilde{h}\widetilde{h}^{\rm T} \ell^{\rm c}_{\rm L} + \frac{1}{2}\frac{y}{\Lambda} \Phi^2 \overline{N^{\rm c}_{}}N  + {\rm  h.c.}+ \cdots \;, %%%
\label{eq:L}
\end{eqnarray}
%%%%%%%%%%%%%%%%%%%%%%%%%
where $\Phi$ represents the ultralight DM field, $h$ is the SM Higgs, the active Majorana neutrino mass term $m^{}_{\nu} \equiv \kappa { v}^2/2$ and the Dirac mass $m^{}_{\rm D} = y^{}_{\rm D} {v}/\sqrt{2}$ will be generated after Higgs takes the vacuum expectation value ${ v}/\sqrt{2} \equiv \langle \widetilde{h} \rangle = 174~{\rm GeV}$, and $m^{}_{N}$ is the Majorana mass term of the sterile neutrino. On top of that, the Yukawa interaction  $g \Phi \overline{N^{\rm c}_{}}N$ with a coupling constant $g$ generates a time-varying mass term to the sterile neutrino. 
In addition, there might be a dimension-five effective interaction with a coupling constant $y$ and the cutoff scale $\Lambda$~\cite{Zhao:2017wmo, Anisimov:2006hv,Anisimov:2008gg}, similar to the Weinberg operator generating light neutrino masses.

In general, the Majorana mass terms $m^{}_{\nu}$ and $m^{}_{N}$ can be vanishing by imposing additional symmetries such as lepton number conservation (then the remaining Lagrangian will look similar to the singlet Majoron model~\cite{Chikashige:1980qk,Chikashige:1980ui,Gelmini:1980re,Choi:1991aa,Acker:1992eh,Georgi:1981pg,Schechter:1981cv}).
In the most economical case, one can even generate small neutrino masses, with a minimal interaction form
$  y^{}_{\rm D} \overline{\ell^{}_{\rm L}} \widetilde{h} N +  g \Phi \overline{N^{\rm c}_{}}N$, but the number of sterile neutrinos should be extended to  at least two in order to explain the oscillation data.
Note that even though we assume Majorana neutrinos in this work, the generalization to Dirac neutrinos is not difficult. 
An alternative scenario is the one in which right-handed neutrinos have no vacuum mass ($m_N = 0$) but get a tiny Majorana mass from their coupling to the scalar field. This can give rise to ultralight scalar-induced Pseudo-Dirac neutrinos~\cite{Dev:2022bae}.
The framework explored in this work is different from the earlier discussions about the mass-varying neutrinos, where the tiny neutrino mass is hypothetically generated from the coupling to the dark energy (e.g., acceleron~\cite{Fardon:2003eh}, quintessence~\cite{Brookfield:2005bz}, etc.).
In that case, neutrino masses are constant over any observable laboratory time scales and one cannot distinguish the induced mass from the vacuum mass by looking for the time-varying signals.
Late-time neutrino mass generation can also arise in models having a late-time cosmic phase transition~\cite{Lorenz:2018fzb}, or models having an anomaly due to a gravitational$-\theta$ term~\cite{Dvali:2016uhn}. In fact, Ref.~\cite{Lorenz:2018fzb} analyzed  redshift dependent neutrino masses in the light of recent {\it Planck} data, and reported a preference for models of late-time neutrino masses. This was further analyzed in a follow-up work~\cite{Lorenz:2021alz}, where additional data from Type-Ia supernovae, as well as structure formation was used. More recently, it was demonstrated that a detection of the diffuse supernova neutrino background could shed light on whether neutrino masses turned on at later redshifts~\cite{deGouvea:2022dtw}.

With the renormalizable terms in Eq.~(\ref{eq:L}), the equation of motion of the scalar in our local galaxy is
%%%%%%%%
\begin{eqnarray} \label{eq:partialphi}
& &(\partial^2 + m^2_{\phi})\Phi =  \frac{g}{2} \left(\overline{N^{\rm c}_{}}N +\overline{N}N^{\rm c}_{} \right) \;.
\end{eqnarray}
%%%%%%%%%
where the addition of Hubble dilution term, $3 H  \dot{\Phi}$ with $H$ being the Hubble expansion rate, is necessary if we consider the scalar evolution over cosmological time scales in the expanding Universe.
In the absence of the source term in the right-hand side, the scalar field evolves freely in the Universe after production.
Because the scalar is assumed to be produced coherently
\footnote{The misalignment mechanism serves as such an example, where a complex scalar field $\Phi$ picks up the expectation value associated with a broken global symmetry, resulting in a massless pseudo Nambu-Goldstone boson. The boson mass, which tilts the Mexican hat, can be generated by some phase transition.} and its occupation number is very high, it is more appropriate to consider $\Phi$ as a classical field instead of a quantum state. Neglecting possible spatial variations (arising from structure formation), which are usually suppressed by the DM velocity $v^{}_{\phi} \sim 10^{-3}$ in our Milky Way, the field evolution simply follows,
%%%%%%
\begin{eqnarray}
\Phi(t) = \phi \sin{m^{}_{\phi}t} \;.
\end{eqnarray}
%%%%%%%%%
Here the time $t$ is calibrated such that $\Phi(0)= 0$, and $\phi =  \sqrt{2\rho}/m^{}_{\phi}$ denotes the field strength, which is approximately $ \phi^{}_{\odot}= 2.15 \times 10^{15}~{\rm eV} \cdot (10^{-18}~{\rm eV}/m^{}_{\phi})$ with the dark matter energy density $\rho \approx 0.3~{\rm GeV \cdot cm^{-3}}$ in our local galaxy.
It is worthwhile to setup the magnitude of the coupling $g$ by noting that $g\phi^{}_{\odot} \approx 2.15~{\rm eV} \cdot (g/ 10^{-15}) \cdot (10^{-18}~{\rm eV}/m^{}_{\phi})$.
Unless otherwise specified, we use the capital $\Phi$ to denote the complete time-varying field and $\phi$ for its amplitude. 

Through the Yukawa coupling to $\Phi$, the sterile neutrino develops an effective time-varying mass term. One may worry about energy-momentum conservation within such a setup. Let us consider a closed system formed only by the scalar and the sterile neutrino. The total Lagrangian explicitly possesses a temporal translation symmetry, so the energy of the entire system must be conserved according to Noether's theorem.  Furthermore, because the Lagrangian with only the neutrino part  preserves spatial translation symmetry, the neutrino momentum must be invariant under the time-varying scalar potential. But the neutrino energy might be perturbed by the scalar field. When the neutrino energy evolves with $E^{2}_{N} = \sqrt{p^{2}_{N} + (g\phi \sin m^{}_{\phi}t)^2}$, the energy of the scalar system $E^{}_{\phi}$ will also change accordingly due to the feedback effect from $N$ source term as in Eq.~(\ref{eq:partialphi}), such that $E^{}_{N} + E^{}_{\phi} = {\rm const}$.

In the absence of the scalar potential, diagonalization of Eq.~(\ref{eq:L}) leads to light mass eigenstates $\nu^{}_{i}$ (for $i = 1,2,3$) and a heavy one $\nu^{}_{4}$.
If the vacuum mass $m^{}_{4}$ is too large compared to the potential $g\phi$, only the effective coupling  by mixing $g^{i j}_{\nu} \Phi\overline{\nu}^{}_{i}\nu^{}_{j} $ will be relevant at low energies, with $i$ and $j$ being the indices of light neutrino mass eigenstates.
However, as has been realized, such a scenario faces inevitable constraints from the observation of cosmic microwave background (CMB)~\cite{Berlin:2016woy,Brdar:2017kbt,Krnjaic:2017zlz,Huang:2021kam}.
The growth of $\phi$ with the redshift will render cosmic neutrinos non-relativistic in the early Universe, which is in contradiction with the free-streaming property of relativistic neutrinos from the CMB observation. Sensitivities of most of the laboratory searches are not even comparable to the CMB constraint in the order of magnitude.

The situation is different in the other regime, when the sterile neutrino mass $m^{}_{4}$ is smaller than $g\phi$. 
In this case, as we go back to the dense early Universe ($ g\phi\gg m^{}_{4}$), the potential induced for light neutrinos $\nu^{}_{i}$ will be suppressed by $ m^2_{\rm D} /(g \phi)$. 
This is very similar to the seesaw mechanism which naturally generates the small neutrino masses with the suppression of heavy degrees of freedom. The details for this observation are given in Appendix~\ref{app:A}. 
The sterile neutrino can be as light as $\mathcal{O}({\rm eV})$, which is relevant for short baseline anomalies~\cite{Gariazzo:2015rra,Giunti:2019aiy,Diaz:2019fwt,Boser:2019rta,Gariazzo:2021wsx,Archidiacono:2022ich}.
A general issue with the light sterile neutrino is the possibility of its thermalization in the early Universe, leading to an increase in $\Delta N^{}_{\rm eff}$ ~\cite{Barbieri:1990vx,Enqvist:1990ad,McKellar:1992ja,Sigl:1993ctk}. In fact, the parameter space suggested by the short-baseline anomalies is almost ruled out by the $\Delta N^{}_{\rm eff}$ constraints from Big Bang nucleosynthesis (BBN) and CMB~\cite{Hannestad:2012ky,Gariazzo:2019gyi}.
As has also been noticed previously~\cite{Zhao:2017wmo,Farzan:2019yvo}, the framework with a time-varying DM field coupled to sterile neutrino can provide a bonus to suppress the production of light sterile neutrinos in the early Universe.

\begin{figure}[t!]
	\centering
	\includegraphics[width=0.79\columnwidth]{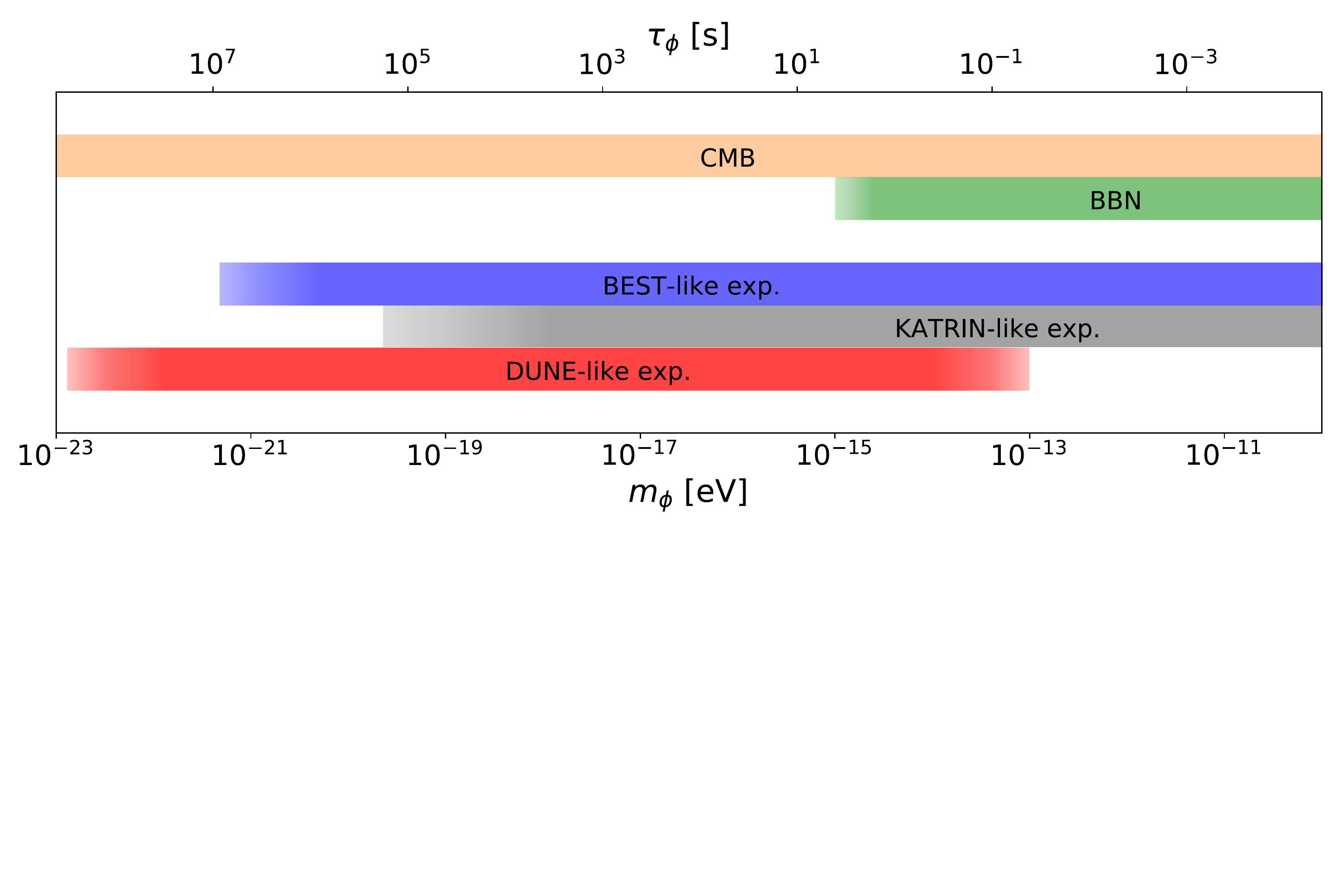}
	\vspace{-4cm}
	\caption{The scalar mass scale relevant for various probes including DUNE-like, KATRIN-like, and BEST-like experiments. The corresponding time scales are indicated on the top axis. The mass ranges for which $m^{}_{\phi} > H$ at $T^{}_{\gamma} = 1~{\rm MeV}$ (for BBN) and $T^{}_{\gamma} = 0.3~{\rm eV}$ (for CMB) are also given for comparison.}
	\label{fig:mass-time}
\end{figure}

In this work, we systematically explore the consequences of a time-varying mass of sterile neutrinos in the early Universe, outlining the impact on BBN and CMB. Furthermore, we discuss the possibility of current generation beta decay experiments like KATRIN to probe the time-varying sterile neutrino hypothesis. Finally, we also highlight how short-baseline neutrino experiments fare in  light of such a hypothesis.
The mass range of the scalar relevant for various probes is illustrated in Fig.~\ref{fig:mass-time}, including the neutrino oscillation experiment DUNE, the beta-decay experiment KATRIN, the short-baseline experiment BEST, as well as the equivalence time scales in the early Universe. The corresponding time scales are indicated on the top axis. It is worth mentioning that black hole superradiance constraints exists in several mass ranges for the ultralight scalar mass \cite{Davoudiasl:2019nlo, Brito:2015oca}.

The structure of the rest of the work is as follows. 
In Sec.~\ref{sec:II}, we separate the time-varying masses via the sterile neutrino portal into two different scenarios. In Sec.~\ref{sec:III}, we investigate the cosmological consequences if the ultralight DM interacts with the sterile neutrino, including both heavy and light sterile neutrino scenarios.
In Sec.~\ref{sec:IV}, we explore the distortion effect induced by the time-varying potential in beta-decay spectrum, taking KATRIN experiment as an example. 
We proceed in Sec.~\ref{sec:V} to discuss the impact of a time-varying light sterile neutrino on the short-baseline experiments. Finally, we discuss our results, and conclude in Sec.~~\ref{sec:VI}.
%%%%%%%%%
%%%%%%%%%
\section{Generating time-varying active neutrino masses from sterile neutrinos}
\label{sec:II}
\noindent
%%%%%%%%%%%%
%%%%%%%%%%%%
\noindent
In the limit that the sterile neutrino is very heavy compared to the scalar potential, i.e., $m^{}_{4} \gg g \phi$, the light neutrinos will receive an effective mass in addition to the original vacuum one (see Appendix~\ref{app:A}),
\begin{eqnarray}\label{eq:mtphi}
\widetilde{m}^{}_{i}(t)  & \approx & 
{m}^{}_{i} +  \sin^2\theta\,  g  {\phi}^{}_{}  \sin{m^{}_{\phi}t} \;.
\end{eqnarray}
The active-sterile mixing angle ${\theta}$ can simply be absorbed by redefining $g^{}_{\nu} \equiv \sin^2{\theta} \cdot g $ such that the neutrino mass correction reads as $g^{}_{\nu} {\phi}^{}_{}  \sin{m^{}_{\phi}t}$. In such a case, it is technically indistinguishable whether active neutrinos couple directly to the scalar field or by mixing with the sterile neutrino, and the discussion will be reduced to the usual scenario explored in the literature~\cite{Berlin:2016woy,Brdar:2017kbt,Krnjaic:2017zlz,Liao:2018byh,Capozzi:2018bps,Reynoso:2016hjr,Huang:2018cwo,Pandey:2018wvh,Farzan:2018pnk,Choi:2019ixb,Baek:2019wdn,Choi:2019zxy,Choi:2020ydp,Dev:2020kgz,Baek:2020ovw,Losada:2021bxx,Smirnov:2021zgn,Alonso-Alvarez:2021pgy,Huang:2021zzz,Huang:2021kam,Chun:2021ief,Reynoso:2022vrn}.

This potential is highly testable in neutrino oscillation experiments, provided the DM oscillation cycle, whose period is given by $2\pi/m^{}_{\phi}$, is at least of the same order as the neutrino time of flight (i.e., the DM field is not \emph{rapidly oscillating}). If the DM field is oscillating with a period of the order of the experimental running time, i.e. few days to years, the neutrino mass eigenstates can develop oscillation phases continuously with constant effective masses during the flight~\cite{Berlin:2016woy,Krnjaic:2017zlz, Dev:2020kgz}. This would manifest as a time modulation of the signal. For shorter DM cycles, the modulation effect will be averaged, but not vanishing. This averaged effect imprints non-trivial distortions to the original neutrino oscillation probability as a function of the neutrino energy. Moreover, it was recently pointed out that even in the rapidly oscillating (dynamical and beyond) regime, the neutrino flavor transition induced by the scalar field is still possible but with a decreased impact~\cite{Dev:2020kgz}. 
\begin{figure}[t!]
	\centering
	\includegraphics[width=0.45\columnwidth]{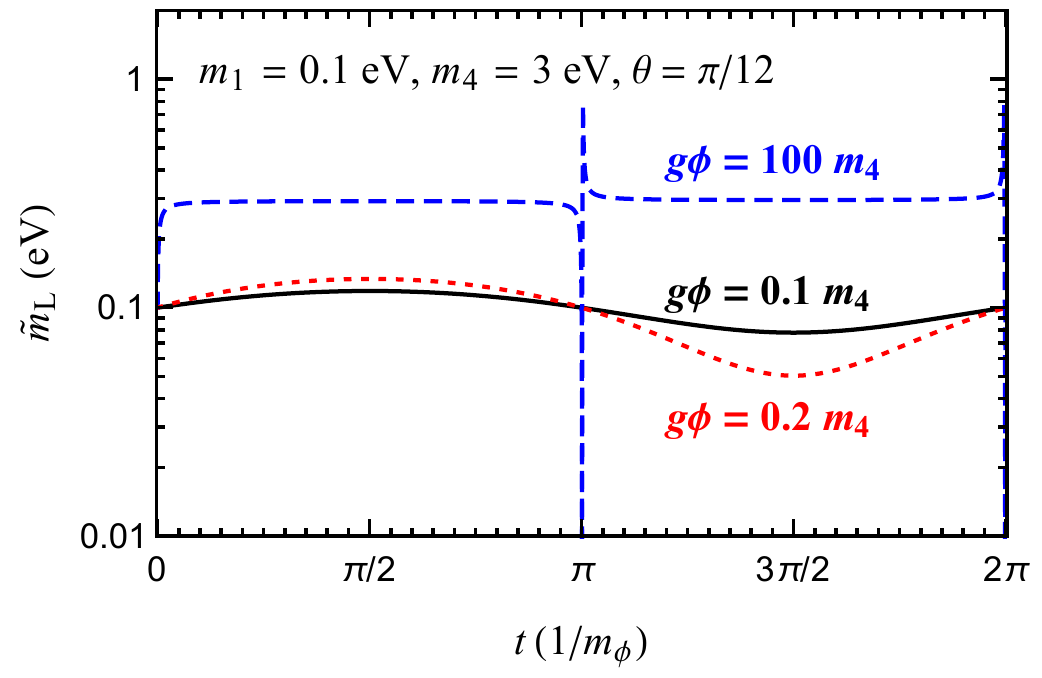}
	\caption{The time-varying neutrino mass $\widetilde{m}^{}_{\rm L}$ via the sterile neutrino portal as a function of time $t$. The vacuum parameters have been chosen as $m^{}_{1} = 0.1~{\rm eV}$, $m^{}_{4} = 3~{\rm eV}$ and $\sin^2{2\theta} = 0.25$. The value of the potential, $g\phi$, has been taken to be $g\phi = 0.1\, m^{}_{4} $ (black curve), $0.2\, m^{}_{4} $ (dotted red curve) and $100\, m^{}_{4} $ (dashed blue curve), respectively. }
	\label{fig:m1_t}
\end{figure}

Even though the time-varying analysis of neutrino oscillation experiments itself is interesting and very rich in phenomenology, the available model parameters are severely constrained from cosmology. 
In fact, a reasonable cosmological scenario, without fine-tuning the DM evolution, might rule out all the parameter space to which neutrino oscillations are sensitive.  
Irrespective of whether the scalar field accommodates all the DM abundance or not, its field strength averaged over space in the early Universe as a function of redshift $z$ will read as
\begin{eqnarray}\label{eq:phiz}
{\phi}(z)^2 = {\phi}(0)^2 \cdot (1+z)^{3} \;,
\end{eqnarray}
up to the time when the Hubble expansion rate becomes comparable to the scalar mass, i.e., $H \approx m^{}_{\phi}$. To see how an experimentally testable coupling is disfavored by cosmology,
we take $g^{}_{\nu} \phi^{}_{\odot} = \sqrt{\Delta m^{2}_{21}} \approx 8.7 \times 10^{-3}~{\rm eV}$, where the local DM overdensity is approximately ${\phi}^{2}_{\rm \odot} \approx 10^5 \, {\phi}(0)^2$ ~\cite{Krnjaic:2017zlz,Zhao:2017wmo}. During the era of matter-radiation equality $(z^{}_{\rm eq} \approx 3000)$, we obtain $g^{}_{\nu} \phi (z^{}_{\rm eq})= 4.5~{\rm eV}$, which  exceeds the neutrino temperature  at that time $T^{}_{\nu} \approx 0.5~{\rm eV}$ by almost one order of magnitude. As a result, neutrinos become nonrelativistic, and do not free-stream at the speed of light before recombination, thereby spoiling CMB observations. The above conclusion is actually rather conservative and requires DM particles to be populated just before the matter-radiation equality. On the other hand, if the DM production is not fine-tuned and takes place early before BBN at $T^{}_{\nu} \approx  1~{\rm MeV}$ (corresponding to $z^{}_{\rm BBN} \approx 6\times 10^9$) such as the misalignment production of QCD axions, a severe limit $g^{}_{\nu}\phi^{}_{\odot} < 7 \times 10^{-7}~{\rm eV}$ can be obtained by conservatively requiring $g^{}_{\nu} \phi( z^{}_{\rm BBN}) < 1~{\rm MeV}$. The allowed tiny local effective time-varying mass clearly rules out all the possibility of realistic laboratory searches.

The above picture changes if $m^{}_{4} < g\phi(z)$ during  the photon decoupling and/or neutrino decoupling era.
In Fig.~\ref{fig:m1_t}, we show the lighter effective neutrino mass $\widetilde{m}^{}_{\rm L} \equiv {\rm Min} \{\widetilde{m}^{}_{1}, \widetilde{m}^{}_{4}\}$  within a DM oscillation period for different $g\phi$.
It is important to note that when the  $g\phi\sin{m^{}_{\phi}t} = - m^{}_{4} - m^{}_{1} $ is satisfied, it is possible for $\widetilde{m}^{}_{\rm L}$ to swap from $\widetilde{m}^{}_{1}$ to $\widetilde{m}^{}_{4}$ due to the resonance encountered. We choose the lighter eigenstate $m^{}_{\rm L}$, because the lighter neutrino is defined to have a dominant overlap with active neutrino, providing the active-sterile mixing is small. 
For the case of $g\phi = 100\, m^{}_{4}$, the lighter mass $\widetilde{m}^{}_{\rm L}$ is given by $\widetilde{m}^{}_{1}$ for $g \Phi > - m^{}_{1} - m^{}_{4}$ and $\widetilde{m}^{}_{4}$ for  $g \Phi < - m^{}_{1} - m^{}_{4}$.
Previous analyses simply assume an ad hoc sinusoidal variation, e.g., Eq.~(\ref{eq:mtphi}), in the active neutrino mass or mixing parameters, which need not necessarily to be the case. As shown in Fig.~\ref{fig:m1_t} and is clear from Eq.~(\ref{eq:wtm1}), the variation of light neutrino parameters can have a more complicated form, especially when $g\phi$ is large compared to the sterile neutrino mass.
In the limit of $0~{\rm eV}<m^{}_{4} \ll g \Phi $, the mass-variation is instead given by (see Appendix~\ref{app:A} for a derivation)
\begin{eqnarray}\label{eq:activemassvariation}
\widetilde{m}^{}_{1}  &\simeq& \frac{m^{}_{1}+m^{}_{4}  - (m^{}_{4} -m^{}_{1})\cos{2\theta}}{2} 
-\frac{(m^{}_{4} -m^{}_{1})^2 \sin^2{2\theta}}{4 g \Phi} \;,\\
\widetilde{m}^{}_{4} &\simeq& \frac{m^{}_{1}+m^{}_{4} +(m^{}_{4} -m^{}_{1})\cos{2\theta}}{2}+ g\Phi\;.
\end{eqnarray}
%%%%%%%%
In the extreme case of $m^{}_{4} \ll g \Phi$ and $\sin{\theta} \ll 1$, a constant shift  $\sim m^{}_{4}  \sin^2{\theta}$ with a time variation with magnitude $ \sim m^{2}_4 \sin^2{\theta}/(g \Phi) $ in the neutrino mass will be induced. 
In this case, we do not expect a large effective neutrino mass in the very early Universe as long as $m^{}_{4}  \sin^2{\theta}$ is chosen to be small.
\begin{figure}[t!]
	\centering
	\includegraphics[width=0.45\columnwidth]{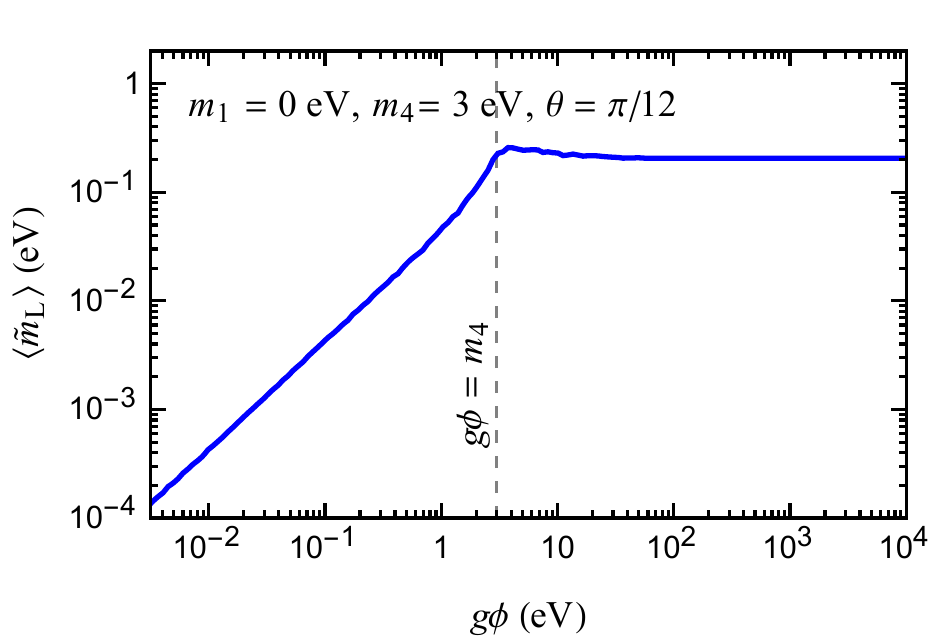}
	\caption{The lighter neutrino mass average $\left\langle \widetilde{m}^{}_{\rm L} \right\rangle$ 
		as a function of the scalar potential, $g\phi$. The vacuum parameters have been chosen as $m^{}_{1} = 0~{\rm eV}$, $m_{4} = 3~{\rm eV}$ and $\sin^2{2\theta} = 0.25$. The turning point around $g\phi = m^{}_{4} $ is marked as the dashed vertical line.}
	\label{fig:m1_gphi}
\end{figure}

\section{Cosmological consequences of time-varying sterile neutrino masses} \label{sec:III}
\subsection{Neutrino decoupling and Big Bang nucleosynthesis}
\noindent
The measurements of primordial helium and deuterium abundances agree very well with the theoretical predictions, assuming standard electroweak interactions with massless neutrinos.  The weak interaction rates are expected to be altered if neutrinos are very massive at the redshift of decoupling, e.g., if $\widetilde{m}^{}_{i} (z^{}_{\rm BBN}) \sim T^{}_{\nu}$. A severe constraint can be imposed on $g \phi^{}_{\odot}$ if ultralight DM particles are assumed to be populated before the BBN era. This can translate into a constraint on the effective neutrino mass for increasing values of 
$g \phi^{}_{}$ as described by Eq.~(\ref{eq:phiz}).

Fig.~\ref{fig:m1_gphi} shows the average of the lighter neutrino mass $\left\langle \widetilde{m}^{}_{\rm L} \right\rangle$ as a function of the potential $g\phi$. 
The vacuum mass and mixing angle have been taken as $m^{}_{1} = 0~{\rm eV}$, $m_{4} = 3~{\rm eV}$ and $\theta = \pi/12$. We observe that starting from very small field strength with $g\phi < m^{}_{4} $, the average mass  $\left\langle \widetilde{m}^{}_{\rm L} \right\rangle$  increases linearly with $g\phi$. At the point around $g\phi = m^{}_{4} $, the average mass $\left\langle \widetilde{m}^{}_{\rm L} \right\rangle$ stops growing as expected from Eq.~(\ref{eq:wtm1_3}).  The turning point around $g\phi = m^{}_{4} $ is the key to alleviate the tension between testable time-varying signals and BBN.The maximum of $\left\langle \widetilde{m}^{}_{\rm L} \right\rangle$, which is given by $\left\langle \widetilde{m}^{}_{\rm L}  \right\rangle_{\rm max} \approx  m^{}_{4}  \sin^2{\theta}$, is under control no matter how $g\phi$ changes in the early Universe.
Hence, in order not to spoil the BBN observations with a large potential, the sterile neutrino parameter should stay within the range $m^{}_{4}  \sin^2{\theta} \ll 1~{\rm MeV}$.

Light sterile neutrinos (lower than the neutrino decoupling temperature, i.e., $m^{}_{4} <1~{\rm MeV}$) have an additional risk of thermalization and increasing the amount of extra radiation in the early Universe, measured by $\Delta N^{}_{\rm eff}$~\cite{Barbieri:1989ti,Dolgov:2003sg}.
It has been noticed that such a risk can be evaded by introducing secret interactions among sterile neutrinos. An effective potential suppressing the active-sterile mixing will hence be induced in the presence of just a small background of sterile neutrinos~\cite{Dasgupta:2013zpn,Hannestad:2013ana,Mirizzi:2014ama,Forastieri:2017oma, Archidiacono:2020yey}.
The spirit is similar in our scenario.
In the presence of a DM potential, sterile neutrino production will be strongly suppressed.
In the following, we numerically investigate this possibility by directly solving for $\Delta N^{}_{\rm eff}$.

\begin{figure}[t!]
	\centering
	\includegraphics[width=0.45\columnwidth]{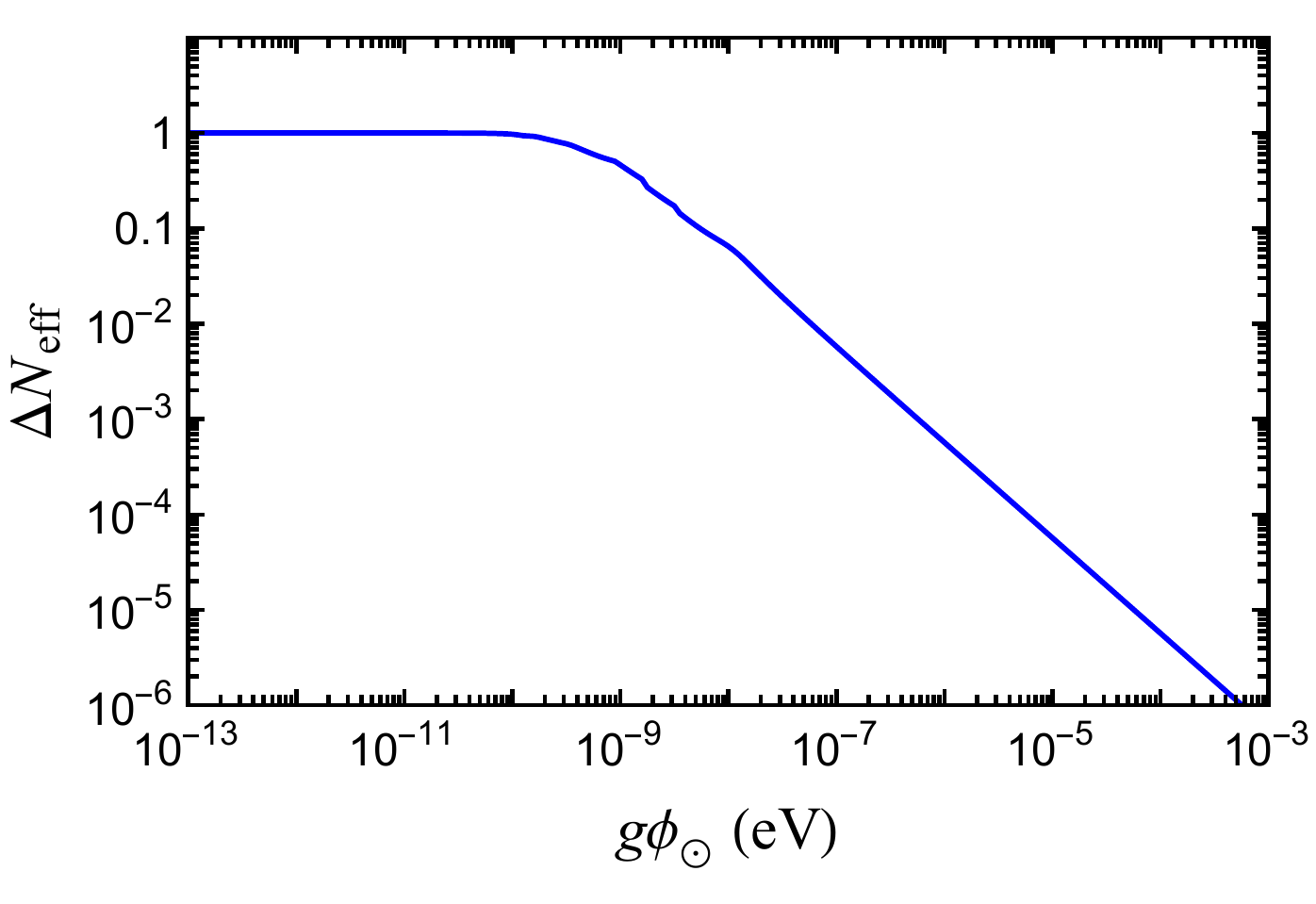}
	\caption{Contribution of sterile neutrinos to the effective number of neutrinos, $\Delta N_{\rm eff}$, around the time of BBN as a function of $g\phi$. The mass splitting and mixing for the sterile neutrino are chosen to be $\Delta m^2_{41} = 10~{ \rm eV}^2$ and $\sin^22\theta = 0.25$.}
	\label{fig:DNeff_gphi}
\end{figure}

In a simplified two-neutrino setup, the evolution equation of the sterile neutrino phase-space distribution $f_{N}$ is given by~\cite{Dodelson:1993je}
\begin{eqnarray}
\label{eq:masterequation}
\frac{\mathrm{d} f^{}_{N}}{\mathrm{d} T^{}_{\nu}} & =- & \frac{\Gamma}{4H T^{}_{\nu}} \,\sin^2\,2 \widetilde{\theta}^{}_{\rm m} \, (f^{}_{\nu_{\rm a}}-f_{N}) \ .
\end{eqnarray}
where $T^{}_{\nu}$ is the neutrino temperature, $f^{}_{\nu_{\rm a}}$ is the active neutrino phase-space distribution, and $\Gamma \propto T^5_\nu/ m_W^4$ encodes the net neutrino interaction rate. Here, $\widetilde{\theta}^{}_{\rm m}$ is the effective mixing angle in matter,
\begin{equation}
\sin^2\,2 \widetilde{\theta}^{}_{\rm m}=   \frac{\langle\widetilde{\Delta}^2\rangle \langle \sin^2\,2\widetilde{\theta}_{14} \rangle}{\langle\widetilde{\Delta}^2\rangle \langle\sin^2\,2 \widetilde{\theta}_{14} \rangle+ \Gamma^2/4 + \left(\langle\widetilde{\Delta}\rangle\langle\cos2 \widetilde{\theta}_{14}\rangle - V_T \right)^2}\,,
\end{equation}
where $\widetilde{\Delta}= (\widetilde{m}^{2}_{4}-\widetilde{m}^{2}_{1})/(2E)$ gives the vacuum oscillation frequency arising from the mass difference between active and sterile species, and $\langle \rangle$ denotes the time-averaging operation. Here $V_T \propto  (T^5_\nu/ m_W^4)$ is a measure of the forward scattering thermal potential experienced by the neutrinos. The scattering term in the denominator encodes the Quantum Zeno effect, where the flavor conversion is suppressed for a large scattering rate.
Furthermore, the self-scattering process $2N  \to 2N$ may freeze-in and become relevant post BBN, leading to a scattering-induced decoherent population of sterile neutrinos~\cite{Mirizzi:2014ama}. However, note that this effect is proportional to the coupling $g$, and is sub-dominant in our case with an extremely small coupling.

In the limit where $f_{\nu_{\rm a}}$ can be approximated by a Fermi-Dirac function (or any generic function of $p/T^{}_{\nu}$), Eq.~(\ref{eq:masterequation}) can be solved approximately to obtain the contribution of sterile neutrinos around the time of BBN~\cite{Jacques:2013xr},
\begin{equation}
\Delta N^{\rm BBN}_{\rm eff}=\frac{f_N}{f_{\nu_{\rm a}}}\simeq 1-{\rm exp}\left[-\frac{2\times 10^3}{4\sqrt{g^*}}\langle \sin^2\,2\widetilde{\theta}_{14} \rangle \frac{\langle\widetilde{m}^{}_{4} \rangle}{\rm eV}\right]\, , 
\end{equation}
where $g^*$ is the effective number of relativistic degrees of freedom. In Fig.~\ref{fig:DNeff_gphi}, we illustrate the dependence of $\Delta N^{}_{\rm eff}$ on the local DM potential $g \phi^{}_{\odot}$ for a given sterile neutrino parameter choice $\Delta m^2_{41} = 9~{\rm eV}^2$ and $\sin^2\,2\theta = 0.25$.
Notably, the presence of just a tiny potential, e.g., $g\phi^{}_{\odot} \gtrsim 10^{-7}~{\rm eV}$, is able to reduce $\Delta N^{}_{\rm eff}$ to a negligible level, i.e., $\Delta N^{}_{\rm eff} \lesssim 0.01 $.
Increasing the potential $g\phi^{}_{\odot}$ will further reduce $\Delta N^{}_{\rm eff}$, thereby making this scenario safe from primordial abundance constraints. One might worry if the presence of the light scalar $\Phi$ itself can act as extra radiation, and run into trouble with BBN predictions. This is again prevented in our scenario due to tiny coupling $g$ considered, i.e., according to the relation $g\phi^{}_{\odot} \approx 2.15 \times 10^{-7}~{\rm eV} \cdot (g/ 10^{-22}) \cdot (10^{-18}~{\rm eV}/m^{}_{\phi})$.

Sterile neutrinos with masses in the keV range and produced through such a freeze-in mechanism can also act as possible dark matter candidates, as was pointed out by Dodelson and Widrow~\cite{Dodelson:1993je}. However, such a mechanism is in tension with the non-observation of X-rays originating from the decay of sterile neutrino dark matter~\cite{Dessert:2018qih}. Nonetheless, in a scenario such as ours, the sterile neutrino need not be a DM candidate, or can only be a tiny fraction of the DM density of the Universe. As a result, all the X-ray bounds will be rescaled by the fractional density of sterile neutrinos in the DM energy budget, and hence, not be relevant for our scenario~\cite{Benso:2019jog}. Interestingly, introducing new interactions among the sterile neutrinos, as well as the active neutrinos have been a popular way to relax these X-ray bounds on these models~\cite{Berlin:2016bdv,DeGouvea:2019wpf}. 

\subsection{Cosmic microwave background and large scale structure}
\noindent
After decoupling around $T^{}_{\nu} = 1~{\rm MeV}$,
neutrinos evolve freely without scattering in our scenario.
However, the neutrino masses will keep varying with the background scalar field, and can be possibly large during the recombination epoch.
Meanwhile, the CMB and large scale structure (LSS) observations are very sensitive to the absolute scale of neutrino masses.
In fact, a world-leading constraint on the sum of neutrino masses has been set, i.e., $\Sigma_i m^{}_{i} < 0.12~{\rm eV}$ using the dataset {\it Planck} TT, TE, EE + lowE + lensing + BAO~\cite{Planck:2018vyg}. Recently, stronger limits, $\sum m_\nu < 0.09$ eV at 95\% C.L. were derived \cite{Palanque-Delabrouille:2019iyz,DiValentino:2021hoh}
One of the key effects of large neutrino masses during recombination is to reduce the free-streaming length $\lambda^{}_{\rm fs}$ compared to the massless neutrino case.
Free-streaming neutrinos can damp all the perturbation modes for distances less than $\lambda^{}_{\rm fs}$, which is well consistent with the current cosmological data. Hence, any effect which reduces the neutrino free-streaming length $\lambda^{}_{\rm fs}$ during recombination can be constrained from CMB data.

\begin{figure}[t!]
	\centering
	\includegraphics[width=0.4\columnwidth]{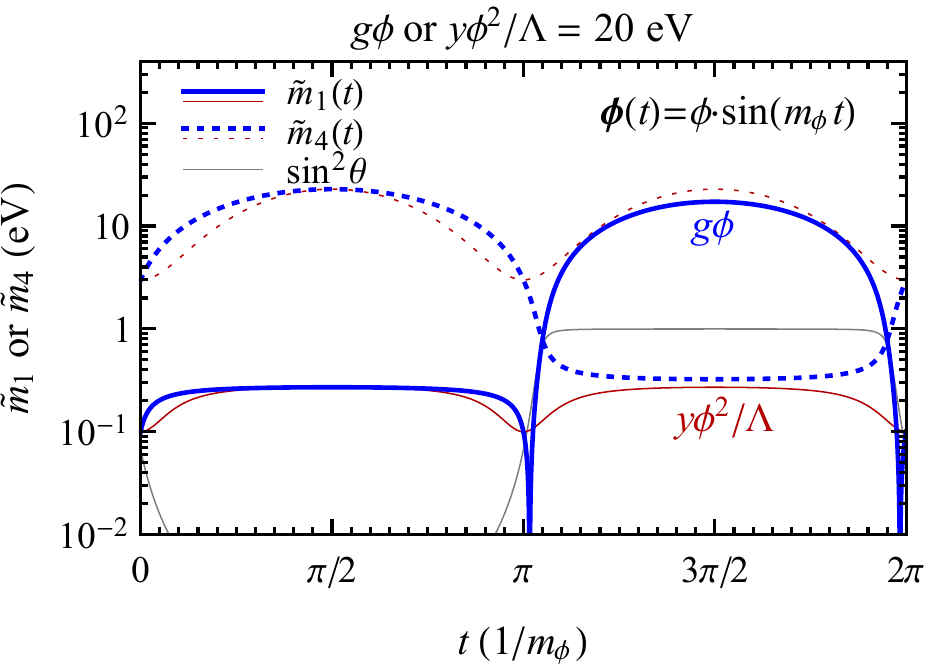}
	\includegraphics[width=0.4\columnwidth]{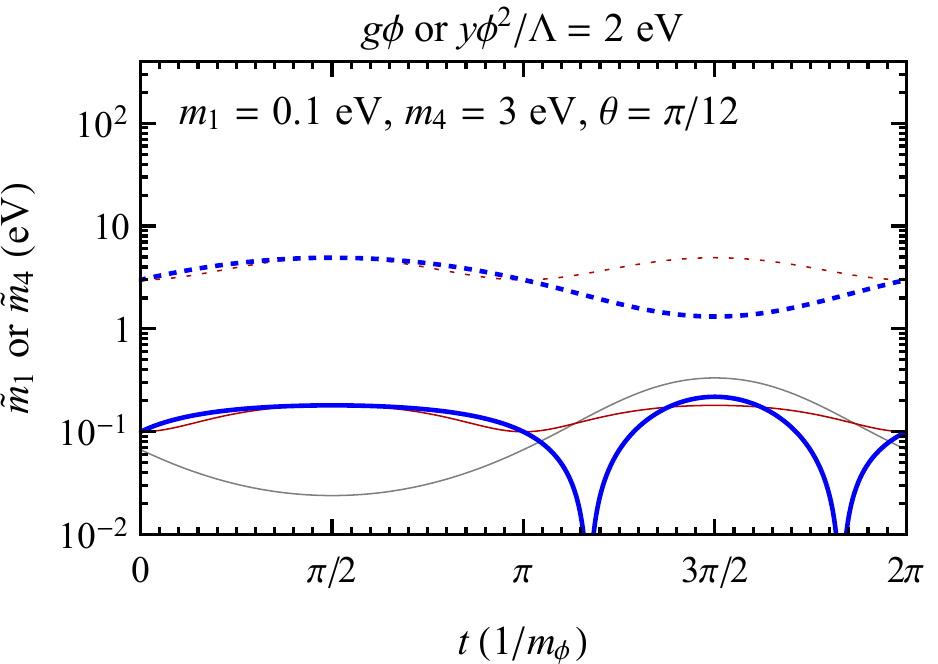}
	\caption{The evolution of $\widetilde{m}^{}_{1}$ (solid curves) and $\widetilde{m}^{}_{4}$ (dotted curves) as functions of time, within one DM cycle. The DM potential is taken to be $g\phi= 20~{\rm eV}$ (blue curves) or $y \phi^2/\Lambda = 20~{\rm eV}$ (red curves) for the left panel. Vacuum neutrino parameters are fixed as $m^{}_{1} = 0.1~{\rm eV}$, $m_{4} = 3~{\rm eV}$ and $\theta = \pi/12$. Conventions are the same for the right panel except that we take $g\phi= 2~{\rm eV}$ or $y \phi^2/\Lambda= 2~{\rm eV}$. }
	\label{fig:m1m4_t1}
\end{figure}

This scenario is different from the BBN epoch, where neutrinos scatter very rapidly before decoupling. In this case, it is important to figure out how a free neutrino mass or flavor eigenstate propagates within the oscillating scalar field. To demonstrate this idea, we plot in Fig.~\ref{fig:m1m4_t1} the two mass eigenvalues $\widetilde{m}^{}_{1}$ and $\widetilde{m}^{}_{4}$ described by Eqs.~(\ref{eq:wtm1}) and (\ref{eq:wtm4}), within one DM oscillation cycle. 
Other parameters are taken as $m^{}_{1} = 0.1~{\rm eV}$, $m^{}_{4} = 3~{\rm eV}$ and $\theta = \pi/12$.
The solid curves stand for the mass $\widetilde{m}^{}_{1}$, which coincides with the vacuum value when $\phi$ is vanishing. On the other hand, the dotted curves are for the $\widetilde{m}^{}_{4}$. 
As $\Phi(t)$ oscillates, we note a non-trivial evolution pattern for $g\phi=20\,$eV (left panel, in blue). At $t = \pi/(2m^{}_{\phi})$, the neutrino masses reach their local maximal value. The lighter neutrino mass is just around $\widetilde{m}^{}_{1} \approx 0.3~{\rm eV}$.
After $t > \pi / m^{}_{\phi}$, $\Phi(t)$ becomes negative, and $\widetilde{m}^{}_{1}$ continuously grows to $\sim 10~{\rm eV}$, which is around the original $\widetilde{m}^{}_{4}$ value. Note that the mixing $\sin^2{\theta}$ (in gray, unitless) also reaches values around one, implying that the corresponding relations of flavor and mass eigenstates are swapped. The other eigenvalue $\widetilde{m}^{}_{4}$ follows an opposite behavior. 

During recombination, if these two mass eigenstates evolve \emph{adiabatically} as $ \Phi(t) $ oscillates, we would expect the neutrino flavors, being a linear combination of the mass eigenstates, to be relativistic half of the time, and non-relativistic the other half. This is equivalent to setting neutrino velocity to $\mathcal{O}(0.5\, c)$, which will hence reduce the free-streaming length $\lambda^{}_{\rm fs}$ by a factor of around two. 
However, we notice that two eigenvalues also critically hit each other at the point of $g \Phi(t) = -m^{}_{4} $, enforcing a non-adiabatic transition. As a consequence, each time a neutrino state, say $\left| \widetilde{\nu}^{}_{1} \right\rangle$, reaches this point, there will be a certain probability for it to transit to $\left| \widetilde{\nu}^{}_{4} \right\rangle$, and vice versa. The net effect is to reduce the neutrino free-streaming length, by at most a factor of two.
We do not investigate numerically the details about the energy-momentum conservation in the scenario where the scalar, active neutrinos and sterile neutrinos form a highly entangled system, but instead give our remarks.
As in the case with only sterile neutrino and scalar, the momentum of neutrinos must be always conserved because of spatial translation symmetry.
When $\left| \widetilde{\nu}^{}_{1} \right\rangle$ transits to $\left| \widetilde{\nu}^{}_{4} \right\rangle$, the energy of each neutrino state is increased by approximately $g\phi$ if $p \ll g\phi$.
However, we notice that $\left| \widetilde{\nu}^{}_{4} \right\rangle$ at $t \gtrsim \pi/m^{}_{\phi}$ has more overlap with the sterile state $\left| N \right\rangle$, hence increasing the contribution of the source term in Eq.~\eqref{eq:partialphi}. This feedback effect should balance the energy between the neutrino and scalar systems.

Even though such a scenario, to our knowledge, has never been strictly investigated with cosmological simulations, one might get an intuition from the neutrino mass limit.
The neutrino mass information one can extract depends on the redshift.
Firstly, the quoted limit from Planck $\sum^{}_{i} m_{i}<0.12~{\rm eV}$ is derived from the data of all available redshifts (from $z=0$ to $z=3000$).
Because the scalar field strength is heavier in the dense early Universe, we are more interested in the consequence at higher redshifts.
There are fits using the CMB and LSS data by allowing neutrino masses to vary with the redshift, which find the mass limit derived from the data at $z>1100$ can only be $\Sigma_{i} m_i < 0.40~{\rm eV}$ at $95\%~{\rm CL}$~\cite{Lorenz:2021alz}.
This can be translated it into a constraint on neutrino velocity, $\left\langle v^{}_{\nu} \right\rangle > 0.97 \, c$, by using the temperature $T^{}_{\nu} (z=1100) \approx 0.18~{\rm eV} $ and the relation $\left\langle p^{}_{\nu} \right\rangle \approx 3 T^{}_{\nu}$.
Even without a dedicated analysis, it indicates that the scenario with $\left\langle v^{}_{\nu} \right\rangle > 0.5 \, c$ is in tension with CMB observations, if $g\phi$ is too large in the early Universe.

However, the above estimate is too stringent, because the impact on CMB perturbation due to a smaller free-streaming is only one of the effects of finite neutrino masses.
The inclusion of more effects, such as background evolution effect, should lead to a more conservative limit on the neutrino velocity, which can only be obtained with a more detailed analysis. Furthermore, it is important to note that while the scalar field keeps oscillating with the frequency $m^{}_{\phi}/(2\pi)$, the overall field strength decreases with Eq.~(\ref{eq:phiz}) as the Universe expands. At certain point when $g\phi < m^{}_{4} $ the two neutrino mass eigenvalues start to separate. This is demonstrated in the right panel of Fig.~\ref{fig:m1m4_t1}, where we show the eigenstates for a smaller value of $g\phi=2\,$eV (blue curve).

The neutrino velocity is reduced in the large $g {\phi}$ case due to the conversion between the lower and upper mass eigenstates when $g \Phi(t) = -m^{}_{4} $.
However, such an issue does not arise if the mass-variation is due to the higher dimensional term, governed by $y {\phi}^2/\Lambda$ case with $y>0$. This is demonstrated via the red curves in Fig.~\ref{fig:m1m4_t1}.
In this case, no matter how large the potential becomes, the eigenstate with $\widetilde{m}^{}_{1}$, which has dominant mixing with active neutrinos, always stays below $m^{}_{4}  \sin^2{\theta}$.
The threats from cosmological observations can thus be removed by this higher-dimensional operator. In such a case, the local potential $y {\phi}^2/\Lambda$  can  be large without spoiling BBN, CMB and LSS to have observable time-varying effect for active neutrinos at laboratories.
Note that our only requirement here is $m^{}_{4}  \sin^2{\theta} < 0.40~{\rm eV}$, where the sterile neutrino mass can actually be heavy, e.g., $m^{}_{4}  = 1~{\rm TeV}$ with $\theta  = 10^{-7}$. 
This requirement is well compatible with the current collider searches of heavy right-handed neutrinos. For instance, the ATLAS and CMS collaborations at LHC have set constraints $|V^{}_{\mu N}|^2, |V^{}_{e N}|^2 \lesssim 10^{-3}$~\cite{ATLAS:2015gtp,CMS:2016aro,CMS:2018iaf} at $m^{}_{4} = 100~{\rm GeV}$, corresponding to a very loose result $m^{}_{4}  \sin^2{\theta} < 0.1~{\rm GeV}$.

In this section, we explored the effects of either $g {\phi}$ or $y {\phi}^2/\Lambda$ term in the early Universe, by keeping the other subdominant. In principle, we can simultaneously have these two effects, where $y {\phi}^2/\Lambda$ is suppressed by some cutoff scale.
However, as we go to the early Universe, the effective term, $y {\phi}^2/\Lambda$, growing faster might dominate over the $g {\phi}$ term, and save the model from cosmological bounds.
In the remaining part of the work, we will explore the consequences of a time-varying neutrino mass on beta decays and light sterile neutrino phenomenology. For these analyses, we shall ignore the higher dimensional operator, and focus on the renormalizable interaction
$g \Phi$ for simplicity.

\section{Tritium beta decays} \label{sec:IV}
\noindent
A time-varying mass of the sterile neutrino can also leave potentially observable imprints in beta-decay experiments, depending on the mass of the sterile neutrino, as well as the mass and amplitude of the scalar field $\Phi$.
When the sterile neutrino is heavy (e.g., $m^{}_{4}  > {\rm MeV}$), larger than $g\phi$ and decoupled from the energy scale of beta-decay experiments, the scalar potential can only affect the beta-decay spectrum by mixing with active ones~\cite{Huang:2021zzz}. In the standard $3\nu$ scenario, beta-decay experiments, such as Mainz~\cite{Kraus:2012he}, Troisk~\cite{Belesev:2012hx, Belesev:2013cba} and KATRIN~\cite{KATRIN:2019yun, KATRIN:2021uub}, measure the effective neutrino mass ${m}^{}_{\beta} \equiv  \sqrt{\sum_{i=1}^3 |{U}^{}_{ei}|^2 m^{2}_{i}}$, which receives incoherent contributions from three generations of neutrinos.
By analyzing the electron spectrum from tritium beta decays, stringent limits have been set, e.g., ${m}^{}_{\beta} < 0.8~{\rm eV}$ from the combination of the first and second campaign of KATRIN (KNM1 + KNM2)~\cite{KATRIN:2021uub}.

In order to show the effect of a time-varying scalar on beta-decay spectrum, we consider a simplified modification to the effective neutrino mass, in the limit $g\phi < m^{}_{4}$, as
\begin{eqnarray}\label{eq:wtmb}
\widetilde{m}^{}_{\beta}(\phi)  & \approx & 
{m}^{}_{\beta} + g^{}_{\nu} {\phi} \sin{m^{}_{\phi}t} \; ,
\label{eqn:heavylimit}
\end{eqnarray}
where $g^{}_{\nu} \equiv \sin^2{\theta}^{}_{14} \cdot g $ represents the effective neutrino coupling suppressed by the active-sterile mixing angle.
For Eq.~(\ref{eq:wtmb}), a uniform mixing of the sterile neutrino  to three generations of neutrinos has been assumed, such that $\widetilde{U}^{}_{ei}(\phi) = {U}^{}_{ei}$ and $\widetilde{m}^{}_{i}(\phi) = {m}^{}_{i}+ g^{}_{\nu}\Phi$ (for $i=1,2,3$) hold, thereby leading to the simplified relation in Eq.~(\ref{eq:wtmb}).
In general, the diagonalization of the sum of vacuum mass matrix and DM-induced mass matrix will result in complex dependence on the scalar field.
For the accuracy of current beta-decay experiments, the major observable is ascribed to a single mass parameter $\widetilde{m}^{}_{\beta}$. Different coupling patterns are assumed to not affect much the overall magnitude of modifications.

The beta spectrum with the effective neutrino mass $\widetilde{m}^{}_{\beta}(\phi) $ can be parameterized as
\begin{equation}\label{eq:betaExact}
R^{}_{\beta}(E^{}_{e},\widetilde{m}^{}_{\beta}) =  \frac{G^2_{\rm F}}{2\pi^3}|V^{}_{\rm ud}|^2  \left(g^{2}_{\rm V} + 3  g^{2}_{\rm A}\right) \frac{m^{}_{^3 {\rm  He}}}{m^{}_{^3 {\rm H}}}  F(Z, E^{}_{e}) \times E^{}_e \sqrt{E^2_{e} - m^2_e}  H(E^{}_{e}, \widetilde{m}^{}_{\beta})   \;,
\end{equation}
with the spectral function
\begin{equation}\label{eq:}
H(E^{}_{e}, \widetilde{m}^{}_{\beta})=
\sqrt{(K^{}_{\rm end,0}-K^{}_{e})^2 - \widetilde{m}^{2}_{\beta} } \times \left(K^{}_{\rm end,0} - K^{}_{e}\right) .
\end{equation}
Here, $G^{}_{\rm F}$ is the Fermi coupling constant, $V^{}_{\rm ud}$ is the weak mixing matrix element, $g^{}_{\rm V}=1$ and $g^{}_{\rm A}=1.247$ are the vector and axial-vector weak coupling constants of Tritium and the function $F(Z,E^{}_{e})$ is the ordinary Fermi function describing the spectral distortion in the atomic Coulomb potential. Throughout this work, we use $E^{}_{e}$ and $K^{}_{e}$ to distinguish the  total and kinematic electron energies and $K^{}_{\rm end,0}$ is the electron endpoint energy in the massless neutrino limit.
The step function, which is necessary to make the square root real, is not explicitly shown.

The ultralight scalar may manifest itself as a modulation effect to the beta spectrum, if the DM cycle can be covered by KATRIN runs, and also be resolvable for the duration of KATRIN spectrum scans.
%%%%%%%%%%%%%%%%
\begin{figure}[t!]
	\centering
	\includegraphics[width=0.45\columnwidth]{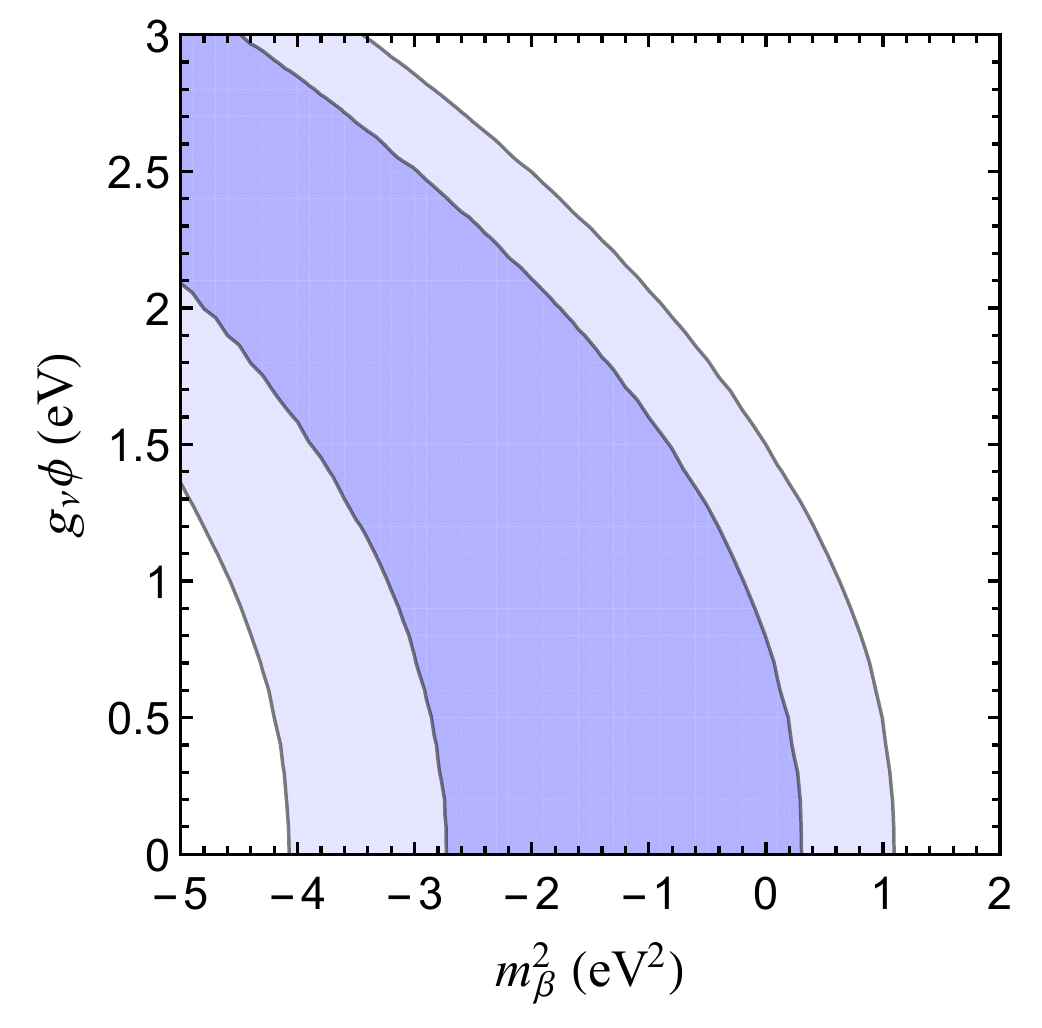}
	\caption{Allowed regions in the ${m}^2_\beta - g_\nu \phi$ plane at 68\% and 95\% C.L. using data from KATRIN's first campaign (KNM1) and in the limit in which $m_4$ is very heavy and $\widetilde{m}^2_\beta$ exhibits a time modulation as in Eq.~\eqref{eq:wtmb}.}
	\label{fig:gphi_mbeta}
\end{figure}
%%%%%%%%%%%%%%%%%%%
If the above condition is not satisfied, one can also look for the distortion effect by averaging over the DM oscillations.
In this circumstance, we investigate analytically how the averaged scalar field modifies the beta spectrum.
For the current KATRIN sensitivity, it is a good approximation to keep up to the first order of perturbative expansions on $\widetilde{m}^{2}_{\beta}$ (unless $g^{}_{\nu}\phi$ is very large) in the spectral function, namely,
\begin{eqnarray}\label{eq:}
H(E^{}_{e}, \widetilde{m}^{}_{\beta}) & \propto &  (K^{}_{\rm end,0}-K^{}_{e}) - \frac{\widetilde{m}^{2}_{\beta}}{2 (K^{}_{\rm end,0}-K^{}_{e})}\; .
\end{eqnarray}
The square of effective neutrino mass as in Eq.~(\ref{eq:wtmb}) averaged over one DM cycle reads
\begin{eqnarray}\label{eq:mbetas}
\left\langle \widetilde{m}^{2}_{\beta} \right\rangle = m^2_{\beta} + \frac{(g^{}_{\nu} \phi )^2}{2} \;.
\label{eqn:effective_mbeta}
\end{eqnarray}
Hence, in such a case the presence of the scalar field directly adds a constant term to the square of effective neutrino mass.
For the sensitivity of first KATRIN campaign, the averaged scalar effect is degenerate with a usual neutrino mass, but it leads to large neutrino mass cosmology~\cite{Alvey:2021xmq}.
This degeneracy is non-linear and obvious from Fig.~ \ref{fig:gphi_mbeta}, where we have fitted the parameter space of $m^2_{\beta}$ and $g\phi$ using the KATRIN data from the first campaign (KMN1). 
Following the analysis strategy from the KATRIN Collaboration, we allow $m^{2}_{\beta}$ to become negative during the fit. From Fig.~\ref{fig:gphi_mbeta} and Eq.~\eqref{eqn:effective_mbeta}, one can see that in this scenario, the effective neutrino mass measured, $\langle \widetilde{m}^{2}_{\beta} \rangle$ is always larger than the true $m^2 _{\beta}$ and hence, the upper limits derived when assuming $g \phi = 0$ are conservative.

Up to this point, we limited the discussion to the case in which the sterile neutrino is heavy. However, when the sterile neutrino is light enough, additional emission channel of beta decays will be open.
In the $(3+1)\nu$ scenario, we can split the $3\nu$ and sterile neutrino contributions as~\cite{KATRIN:2020dpx}
%%%%%%%%%%%%
\begin{equation}\label{eq:}
R^{(3+1)\nu}_{\beta}(E^{}_{e}) = \left(1- \left|\widetilde{U}^{}_{e4}(\phi) \right|^2 \right)	R^{}_{\beta}(E^{}_{e},\widetilde{m}^{}_{\beta}) + \left|\widetilde{U}^{}_{e4}(\phi) \right|^2 R^{}_{\beta}(E^{}_{e},\widetilde{m}^{}_{4}) \;,
\end{equation}
%%%%%%%%%%%%
with $|\widetilde{U}^{}_{e4}|= \sin{\widetilde{\theta}}$. The $\phi(t)$-dependent mixing angle $\widetilde{\theta}$ as well as  masses $\widetilde{m}^{}_{\beta}$ and $\widetilde{m}^{}_{4}$ can be  calculated from Eqs.~(\ref{eq:wtm1}), (\ref{eq:wtm4}) and (\ref{eq:wtth14}), respectively. For simplicity, we assume that the $\phi(t)$ dependence in $\widetilde{m}^{}_{\beta}$ is approximately that of $\widetilde{m}^{}_{1}$. This is well justified in light of the current KATRIN sensitivity to the effective neutrino mass. 

\begin{figure}[t!]
	\centering
	\includegraphics[width=0.4\columnwidth]{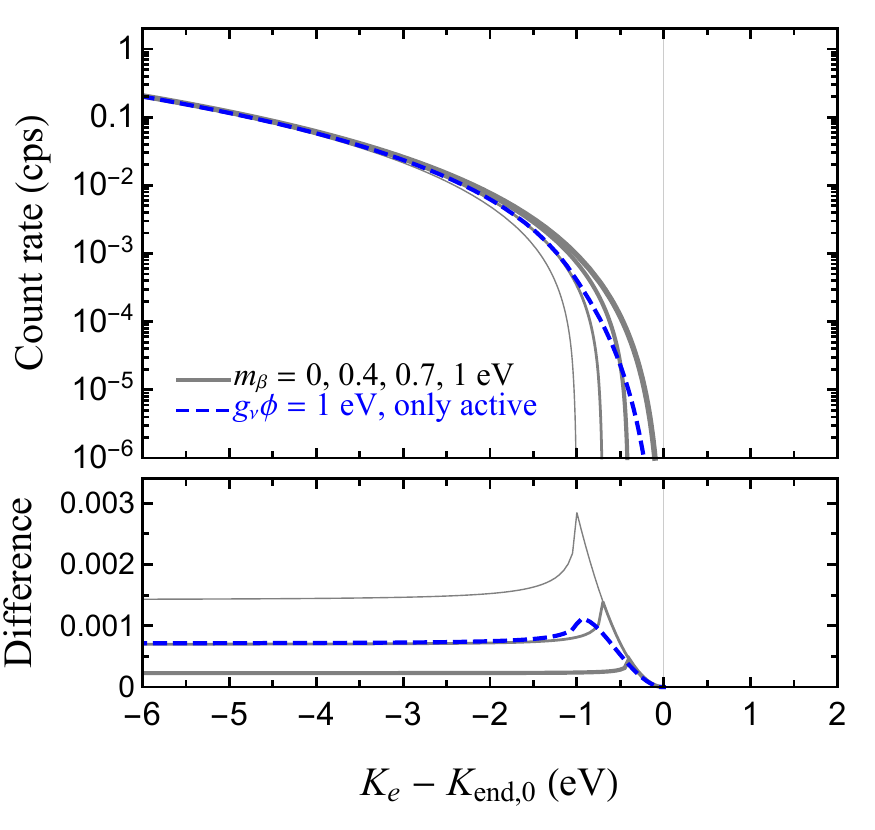}
	\includegraphics[width=0.4\columnwidth]{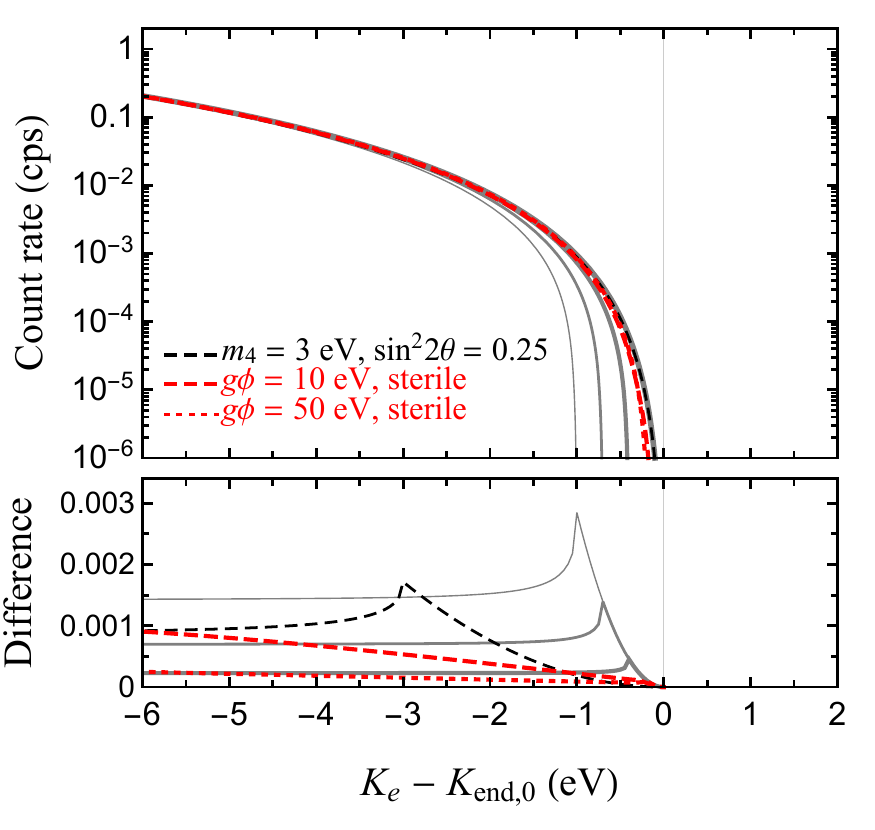}
	\caption{The beta-decay spectra for various scenarios including: the standard beta decays with $m^{}_{\beta} = \{0, 0.4, 0.7,1\}~{\rm eV}$ (gray curves from right to left), the heavy sterile neutrino case as in Eq.~\eqref{eqn:heavylimit} with an effective potential $g^{}_{\nu} \phi = 1~{\rm eV}$ (left panel, blue curve), a benchmark choice of light sterile neutrino with $m^{}_{4}  = 3~{\rm eV}$ and $\sin^2{2\theta} = 0.25$ (right panel, black dashed curve), and the addition of scalar potentials $g\phi = 10~{\rm eV}$ (red dashed curve) and $50~{\rm eV}$ (red dotted curve) upon the light sterile case. Note that we assume $m^{}_{\beta} = 0$ for all the cases with non-zero scalar potentials.}
	\label{fig:spectrum}
\end{figure}

For a general light sterile neutrino, one has to integrate the exact spectral function over DM modulations. 
A generic numerical treatment can be performed in the  analysis  without making approximations.
Aiming to provide a deeper comprehension of the experimental signatures, we show in the upper panel of Fig.~\ref{fig:spectrum} the beta-decay spectra for different scenarios, with a configuration similar to the first KATRIN campaign. The lower panel demonstrates the difference of rates of various scenarios with respect to the standard one with $m^{}_{\beta} = 0~{\rm eV}$.
The region where KATRIN can obtain most information about neutrino masses is slightly above the beta-decay endpoint. The statistics are very limited close to the endpoint, while far above the endpoint the beta-decay rate is too large and hence, insensitive to very small distortions.
A finite neutrino mass will induce not only kink structure, but also provide a constant shift to the beta-decay spectrum away from the endpoint. This is illustrated in Fig.~\ref{fig:spectrum} for four different values of $m^{}_{\beta} = \{0, 0.4, 0.7,1\}~{\rm eV}$, which correspond to the four gray lines, from right to left.
However, the kink structure is still unresolvable given the sensitivity of current generation of beta-decay experiments, resulting in the current limit of $m^2_\beta < 0.9$ eV from KATRIN's second campaign \cite{KATRIN:2021uub}.  

A light sterile neutrino will induce a second kink in the decay spectrum. Such spectral feature, which is expected around $K_e - K_{\rm end,0} \lesssim m_4$, can be well separated from a normal neutrino mass term providing the kink is away from the endpoint, as shown by black dashed curve. The size of the distortion is related to the size of the mixing $|U_{e4}|^2$.  Consequently, beta-decay experiments set limits to the sterile neutrino mass and mixing \cite{Giunti:2019fcj,KATRIN:2022ith}.
Red curves in Fig.~\ref{fig:spectrum} illustrate how adding large scalar potentials to the sterile neutrino will smooth these distortions, mainly because the averaged mixing will be suppressed by the potential. Consequently, this scenario allows to open up the sterile neutrino parameter space in the context of short-baseline anomalies, as we will discuss later in Fig.~\ref{fig:SBL}.
The case in which the sterile neutrino is heavy is given by the the blue curve. We have shown that the effect of the scalar is degenerate with the mass term (see Fig.~\ref{fig:gphi_mbeta}).

\section{Impact on gallium and reactor anomalies}\label{sec:V}
\noindent 
Until now, we have discussed the impact of a consistent scenario of time-varying neutrino masses on cosmology and beta decay experiments. If the sterile neutrino is light today, it will also have important consequences for the long-standing gallium and reactor anomalies. We address these issues in this section. 
First, we will focus on the case where the DM cycle matches the time scale of the experiment, giving rise to a signal of time modulation. We use the latest BEST experiment as a case study. Second, we move to higher scalar masses in the rapidly oscillating regime and discuss its impact on the  parameter space of light sterile neutrinos.
%%%%%%%%%%
%%%%%%%%
\subsection{Time modulation and the gallium anomaly results}

\noindent 
Solar neutrino detectors GALLEX~ \cite{Kaether:2010ag} and SAGE~\cite{SAGE:2009eeu} have used gallium to measure the neutrino emission rate from radioactive ${}^{51}$Cr and ${}^{37}$Ar sources, and found the rates to be lower than the prediction. This is known as the gallium anomaly, which can be, in principle, explained by adding an  eV-mass light sterile neutrino to the SM. This anomaly has been recently strengthened by the  BEST collaboration, which detects $\overline{\nu}_e$ from a ${}^{51}$Cr source kept in a two-volume detector, filled with gallium~\cite{Barinov:2021asz,Barinov:2022wfh}. BEST provides high statistics real-time data of the event rates in the inner as well as outer detectors.
The purpose of this subsection is to illustrate that the type of signal predicted by the time-varying scenario could manifest as anomalous experimental results like the one reported by BEST, and the time modulation probed would correspond to the ultralight scalar mass range of interest. To do so, we performed a simple fit of the available data considering $m_4$, $\theta_{14}$ and $g\phi$ as free parameters.
%%%%
\begin{figure}[t!]
	\centering
	\includegraphics[width=0.4\columnwidth]{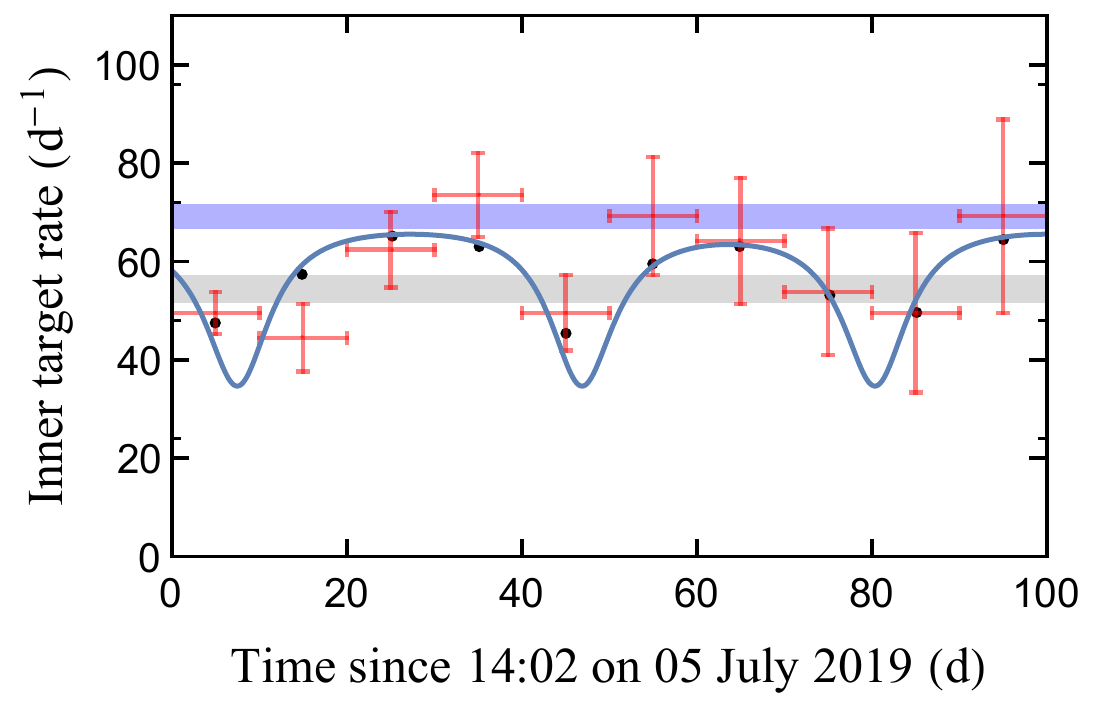}
	\includegraphics[width=0.4\columnwidth]{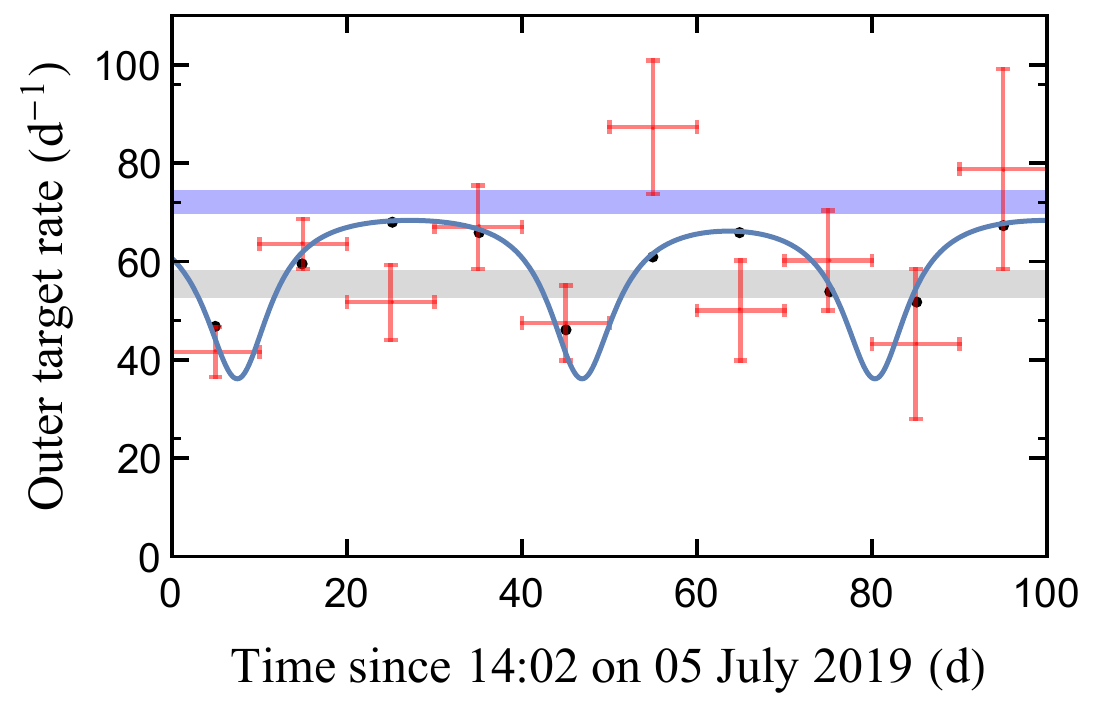}
	\caption{The neutrino reaction rate at BEST for inner (left panel) and outer (right panel) targets, starting from 14:02 on 05 July 2019. The event rates with error bars (in red) are taken from Ref.~\cite{Barinov:2022wfh}. The purple band is the original rate expectation without any deficits, and the gray one is the best-fit scenario with a constant neutrino flux deficit. A sterile neutrino with the time-varying potential generates the blue curves. To be more specific, the parameters are chosen as $m^{}_{4}  = 3~{\rm eV}$, $\theta = \pi/5$, $g\phi = 4.2~{\rm eV}$ and $m^{}_{\phi} = 6.6 \times 10^{-22}~{\rm eV}$.}
	\label{fig:gallium}
\end{figure}

Our results are shown in Fig.\,\ref{fig:gallium}, which depicts the neutrino reaction rate at inner and outer targets, starting from 14:02 on 05 July 2019. 
The event rates with error bars, shown in red, are taken directly from Ref.~\cite{Barinov:2022wfh}.
The purple band is the original prediction without any electron neutrino disappearance.
The gray band represents the best-fit scenario with a constant neutrino flux deficit.
The model with time-varying sterile neutrino mass generates the blue curve, which clearly shows a periodic behaviour in addition to a reduction in the expected $\nu_e$ flux.
To generate this curve, we use the best-fit values, $m^{}_{4}  = 3~{\rm eV}$, $\theta = \pi/5$, $g\phi = 4.2~{\rm eV}$ and a DM mass $m^{}_{\phi} = 6.6 \times 10^{-22}~{\rm eV}$ (corresponding to a cycle of 73 days), which is right in the fuzzy DM regime.  
We find that the original fit with constant flux deficit gives  a global $\chi^2 \approx 32$, while the time-varying scenario can reduce it to $\chi^2 \approx 18$.
Then, it might seem that the time-varying massive sterile neutrino scenario provides a better explanation of the BEST results than the vanilla sterile neutrino case.
Note that the procedure followed did not account for possible time-dependent systematics. Besides that, in order to actually claim the existence of a time modulation in data, a more sophisticated analysis is required in order to address if the periodic behaviour in data is spurious (for instance, due to statistical fluctuations) and its statistical significance.  Finally, in order to actually conclude whether this scenario provides a better fit to the data, one would need to perform a proper Monte-Carlo statistical analysis which allows to interpret consistently the meaning of the lower $\chi^2$ obtained.

Apart from that, the interpretation of the anomaly in terms of a time-varying sterile neutrino is in conflict with solar neutrino data and other active neutrino oscillation experiments.
The active to sterile oscillation will lead to a deficit in the solar neutrino flux as well \cite{Kopp:2013vaa,Dentler:2018sju}.
The original BEST fit is already in tension with the solar neutrino data, which cannot be saved by simply adding a modulation, unless there is a large overdensity of DM inside the Sun (e.g., ten times of that on the Earth).
Furthermore, by mixing, the active neutrino masses will unavoidably receive time-varying corrections.
However, no modulation pattern or averaged effect have been found in existing oscillation experiments~\cite{DayaBay:2018fsh,Borexino:2022khe} so far. In particular, modulations with an amplitude larger than 10\% are excluded by solar data for modulation periods ranging from $\mathcal{O}$(10 minutes) to $\mathcal{O}$(10 years)~\cite{Super-Kamiokande:2003snd,SNO:2005ftm, SNO:2009ktr}. It remains to be seen whether the modulation pattern persists or not in future data collections.
Should the modulation pattern be caused by other uncontrolled time-dependent systematics rather than time-varying sterile neutrinos, we may leave it for a future analysis.
Nevertheless, this serves as an illustrative example of how an experiment like BEST could constrain the time-varying neutrino mass.

%%%%%%%%%%%%%%%%%%
%%%%%%%%%%%%%%%%%%
\begin{figure*}[t]
	\centering
	\includegraphics[width=0.9\columnwidth]{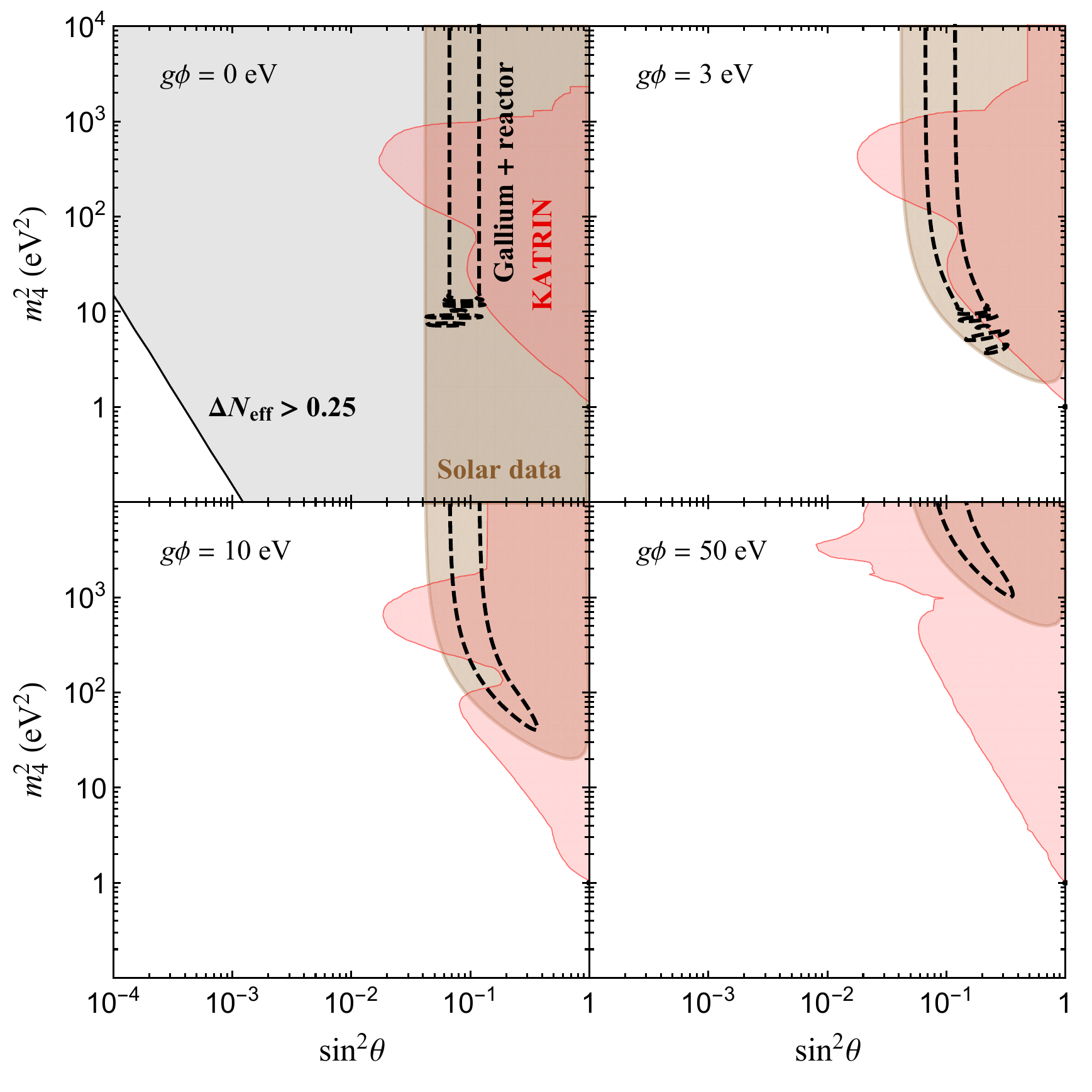}
	\caption{The light sterile neutrino parameter space in the presence of the rapidly oscillating DM potential. From top left to bottom right, we have set $g\phi = 0~{\rm eV}, 3~{\rm eV}, 10~{\rm eV}$ and $50~{\rm eV}$.
		The lightest neutrino is fixed to zero. Black curves enclose the region favored by the global analysis of reactor and gallium data. 
		Several regions are excluded from solar neutrino data (in brown) and KATRIN (in red). The parameter space excluded based on $\Delta N^{}_{\rm eff} < 0.25$~\cite{Archidiacono:2022ich,Planck:2018vyg} from {\it Planck} TT, TE, EE + lowE + lensing + BAO is shown in gray, while for the other three cases the resultant $\Delta N^{}_{\rm eff}$ is vanishingly small.  }
	\label{fig:SBL}
\end{figure*}

%%%%%%%%%%%%
%%%%%%%%%%%%
\subsection{Reinterpretation of light sterile neutrino parameter space}
\noindent
%%%%%%%%%%%%
%%%%%%%%%%%%
The gallium anomaly is further strengthened by the presence of the reactor antineutrino anomaly -- a series of reactor experiments which observed a deficit of the electron antineutrino flux as compared to some predictions (see ~\cite{Giunti:2019aiy, Giunti:2021kab} for recent reviews).
In this subsection, we will discuss how the light sterile neutrino parameter space changes in connection with the gallium as well as the reactor anomalies under the oscillating scalar potential hypothesis. Note that a typical reactor neutrino experiment baseline is small, e.g., baselines are of the order 0.01-1~{\rm km}, and therefore is sensitive to DM modulations driven by a mass (or equivalently frequency) $m^{}_{\phi} \approx  10^{-7} -  10^{-9}~{\rm eV}$.

For simplicity, we will focus on the scenario where the
DM oscillates very rapidly during the neutrino flight from the source to the detector, such that all the time-varying information is effectively lost. Since the DM potential encountered by the neutrinos changes rapidly, neutrinos in flight do not have enough time to develop any non-trivial phases due to this interaction.
The Hamiltonian relevant for active-sterile neutrino oscillations is reduced to a vacuum term and a time-dependent one, namely $H= H^{}_{0} + H^{}_{\rm I}(t)$ with
\begin{eqnarray}
H^{}_{0} & = &  \frac{1}{2p}\left[\begin{matrix}
m^2_{\nu} + m^2_{\rm D} & m^{}_{\rm D}(m^{}_{\nu}+m^{}_{N} ) \\
m^{}_{\rm D}(m^{}_{\nu}+m^{}_{N} ) &  m^2_{\rm D}+ m^{2}_{N}
\end{matrix} \right] ,  \\ 
H^{}_{\rm I}(t) & = &  \frac{1}{2p}\left[\begin{matrix}
0 & m^{}_{\rm D} \, g \Phi \\
m^{}_{\rm D}\, g \Phi & 2 m^{}_{N} g\Phi+(g \Phi)^2
\end{matrix} \right]. \label{eq:HIt}
\end{eqnarray}
In the event of rapid DM oscillations, any terms in the Hamiltonian proportional to single powers of $g{\Phi}$ will be averaged to zero. On the other hand, the quadratic term will be averaged to $(g{\phi})^2/2$. This constant term behaves similar to a matter potential, causing a shift in the dispersion relations; see Appendix~\ref{app:C} for further details.

Fig.~\ref{fig:SBL} shows the impact of a rapidly oscillating DM, coupled to sterile neutrinos, on the different types of constraints existing in the sterile neutrino parameter space for four different benchmark potential values, $g\phi = 0$, $g\phi = 3~{\rm eV}$, $g\phi = 10~{\rm eV}$ and $g\phi = 50~{\rm eV}$ (from top left to bottom right). The red shaded regions indicate our KATRIN limits. The constraint from the $\Delta N^{}_{\rm eff}$  requirement~\cite{Archidiacono:2022ich} is shown as the gray region in the top-left panel, while there are no such constraints for the other three cases. The black dashed lines show the region favored by gallium and reactor data~\cite{Berryman:2021yan}, while the brown shaded region indicate the solar neutrino constraint~\cite{Goldhagen:2021kxe}. For convenience, the active neutrino masses have been fixed to zero in all the cases.
We find that as $g\phi$ increases, the favored parameter space from the global analysis of gallium and reactor data shrink to the top-right corner. This is due to the fact that for $g\phi \gg m^{}_{4} $, the active-sterile mixing will be severely suppressed, and one needs to go to larger vacuum mixings to have the same effective parameters. Similar trends are seen for the solar bounds as well.
Regarding KATRIN, for increasing values of $g\phi$, larger values of the mixing become allowed for masses below 10 eV. The bumpy structure present in the vanilla sterile neutrino constraint (top-left panel), which arises as a consequence of statistical fluctuations, is kept approximately in all cases.
The analysis has been limited to the case of $m^{}_\beta = 0~{\rm eV}$. Including this parameter in the fit should smooth the displayed curves. As a related issue, one can see that for non-zero $g\phi$, maximal mixing is not allowed for sterile neutrino masses of $\mathcal{O}({\rm 50 -100 eV})$. That region of parameter space corresponds to the limit in which sterile neutrinos are integrated out and only the time modulation affecting active neutrinos can manifest in beta-decay experiments. In that case, there is a degeneracy between $g\phi$ and $m^2_\beta$ (see discussion in Section \ref{sec:IV}), so it is likely that in a more general analysis by including $m^2_\beta$ as a free parameter, the constraint in that region will be relaxed.
As discussed before, for neutrino oscillation experiments, only  $(g\phi)^2$ term exists when we average over the rapidly oscillating scalar field.  As a result, the mixing angle suppression goes as  $\widetilde{\theta} \sim \theta \cdot \widetilde{m}^{2}_{4}/(g\phi)^2$.
On the other hand, for beta-decay experiments, the suppression to the mixing angle is much milder $\widetilde{\theta} \sim \theta \cdot {m}^{}_{4}/(g\phi)$ because the oscillating scalar field is not averaged at the Hamiltonian level as in the neutrino oscillation experiment.

There are  additional constraints in the parameters space of light sterile neutrinos coming from oscillation experiments (see Ref.~\cite{Acero:2022wqg} for a recent review and references therein for a detailed description of existing limits). We leave the study of the rich phenomenology expected in that case to a future work.

\section{Conclusion} \label{sec:VI}
\noindent
The ultralight DM, as an alternative DM candidate, can develop large interactions with the invisible neutrinos. Such interactions have triggered studies at various neutrino experiments like oscillations, beta decays and $0\nu\beta\beta$ decays.
However, active neutrinos and charged leptons form doublets in the SM, and hence caution should be taken when we couple neutrinos to the exotic fields.
One of the safest ways is to introduce a SM singlet, e.g., sterile neutrino, which can share the exotic forces with light neutrinos by mixing.

In that spirit, we have systematically explored the time-varying neutrino masses by assuming that the sterile neutrino couples to an ultralight scalar field.
The original time-varying neutrino masses, if testable in neutrino oscillation experiments, are actually in severe tension with the neutrino mass constraints from Big Bang nucleosynthesis, cosmic microwave background, and large scale structures.
We demonstrate that the coupling with light sterile neutrinos can provide time-varying signals in active neutrinos  while being in accordance with cosmological observations. However, to evade the constraints from CMB and LSS, it might be necessary to introduce higher-dimensional operators.
If the sterile neutrino is light, there are additional concerns about producing extra radiation prior to BBN, as measured by $\Delta N^{}_{\rm eff}$. We study this scenario, and show that the BBN constraint on light sterile neutrinos can be safely avoided in this case. Additional thermal production of the light DM is not of concern due to tiny couplings in our scenarios. This allows us to provide a cosmology-friendly scenario of time-varying masses of active neutrinos.

We show that the coupling of the ultralight scalar to neutrinos can manifest at beta-decay experiments like KATRIN. We perform a dedicated analysis of the impact of such time varying neutrino masses in KATRIN. We find that in the presence of additional light sterile neutrinos, the limits set by KATRIN change considerably. 

Furthermore, we also study the impact of such models on light sterile neutrino searches, with emphasis on the results from reactor and gallium experiments, as well as solar data.
We find that as the value of the DM mass increases, the region favoured by gallium and reactor data shrink to larger values of sterile neutrino mass and mixing angles. This is due to the large suppression of active-sterile mixing angle, leading to the necessity of large vacuum mixing angles and masses to satisfy the data. Additionally, in the latest gallium data from the BEST collaboration, we find that the time-varying neutrino mass can give rise to signatures in the direction of the observed anomalous results, with the hypothesis, giving a lower $\chi^2$ value, $\Delta \chi^2 = -14$, with respect to the original fit. However, such a scenario turns out to be in conflict with other neutrino oscillation experiments, pointing to other possible origin of the modulation pattern.

Such a framework of time-varying neutrino masses can not only be tested against cosmological observations, but also provide observable signatures in beta decay experiments like KATRIN. Furthermore, it also holds potent consequences for short baseline anomalies
that plague neutrino physics these days.

\acknowledgments
\noindent We thank Pedro Machado and Yue Zhang for insightful comments on the draft and Sergio Pastor for useful discussion. PMM is grateful for the kind hospitality of Particle and Astroparticle Division of the Max-Plank-Institut für Kernphysik (Heidelberg) during the early stages of this work. GYH is supported by the Alexander von Humboldt Foundation. PMM is supported by FPU18/04571 and the Spanish grant PID2020-113775GB-I00 (AEI / 10.13039/501100011033).

\appendix

\section{Effective neutrino masses and mixing with a time-varying scalar}\label{app:A}
\noindent
In the two-flavor (active $\nu^{}_{\rm a}$ and sterile $N$) frame, starting from Eq.~\eqref{eq:L}, we derive the effective neutrino masses and mixings in the presence of the scalar potential $g\Phi$. In the absence of $g\Phi$, the diagonalization of the mass matrices leads to a vacuum mixing matrix $U$, and two masses $m^{}_{1}$ and $m^{}_{4}$, which dominantly mix with active and sterile neutrinos, respectively.
The addition of the scalar potential leads to a mass matrix in the flavor basis $(\nu^{}_{\rm a}, N^{\rm c}_{})$ as
\begin{eqnarray}\label{eq:}
\widetilde{M}^{}_{\nu} & = &  U^{\dagger}
\left( \begin{matrix}
m^{}_{1} & 0 \\
0 & m^{}_{4}  
\end{matrix} \right)  U^* 
+ \left( \begin{matrix}
0 & 0 \\
0 & g^{}_{} \Phi
\end{matrix} \right) \notag\\ 
& = & \widetilde{U}^{\dagger}
\left( \begin{matrix}
\widetilde{m}^{}_{1} & 0 \\
0 & \widetilde{m}^{}_{4} 
\end{matrix} \right)  \widetilde{U}^*  \;,
\end{eqnarray}
where the eigenvalues satisfy the sum rule $\widetilde{m}^{}_{1}+\widetilde{m}^{}_{4} = {m}^{}_{1}+{m}^{}_{4}+ g^{}_{}\Phi$.
The effective neutrino mass and mixing in the presence of ultralight scalar have the following strict expressions
\begin{eqnarray} \label{eq:wtm1}
& & 2\widetilde{m}^{}_{1} =  m^{}_{1} + m^{}_{4}  + g\Phi  - \left[ (m^{}_{4} -m^{}_{1})^2+(g \Phi)^2  +\, 2 (m^{}_{4} -m^{}_{1}) g\Phi \cos{2 \theta^{}_{14}} \right]^{1/2}\;, \\ \label{eq:wtm4}
& & 2\widetilde{m}^{}_{4} =  m^{}_{1} + m^{}_{4}  + g\Phi  + \left[ (m^{}_{4} -m^{}_{1})^2+(g\Phi)^2 +\, 2 (m^{}_{4} -m^{}_{1}) g\Phi\cos{2 \theta^{}_{14}} \right]^{1/2}\;, \\ \label{eq:wtth14}
& &  \tan{2\widetilde{\theta}^{}_{14}}= \frac{(m^{}_{4} -m^{}_{1}) \sin{2\theta^{}_{14}}}{(m^{}_{4} -m^{}_{1}) \cos{2\theta^{}_{14}} + g\Phi} \;.
\end{eqnarray}
When $|g\Phi| \ll m^{}_{4}  \cos{2\theta^{}_{14}}$, we have the approximation
\begin{eqnarray}\label{eq:wtm1_2}
& &\widetilde{m}^{}_{1} \simeq   m^{}_{1} + \sin^2{\theta^{}_{14}} \cdot  g \Phi  \;, \\
& &\widetilde{m}^{}_{4} \simeq m^{}_{4}  + \cos^2{\theta^{}_{14}} \cdot  g \Phi \;, \\ 
& &\tan{2\widetilde{\theta}^{}_{14}}  \simeq  \tan{2{\theta}^{}_{14}} \;,
\end{eqnarray}
which usually applies to the case with heavy sterile neutrinos.
In the early Universe, we often have $|g \Phi| \gg m^{}_{4} $, which will lead to another useful approximation depending on the sign of $g \Phi$. In the case of $g \Phi \gg m^{}_{4} >0~{\rm eV}$, we have the approximation
\begin{eqnarray}\label{eq:wtm1_3}
& & \widetilde{m}^{}_{1} \simeq \frac{m^{}_{1}+m^{}_{4}  - (m^{}_{4} -m^{}_{1})\cos{2\theta^{}_{14}}}{2} - \frac{(m^{}_{4} -m^{}_{1})^2 \sin^2{2\theta^{}_{14}}}{4 g \Phi} \;, \\
& & \widetilde{m}^{}_{4} \simeq \frac{m^{}_{1}+m^{}_{4} +(m^{}_{4} -m^{}_{1})\cos{2\theta^{}_{14}}}{2}+ g\Phi \;, \\
& & \tan{2\widetilde{\theta}^{}_{14}} \simeq  \frac{(m^{}_{4} -m^{}_{1}) \sin{2\theta^{}_{14}}}{ g\Phi}\;.
\end{eqnarray}
Whereas, in the case of $g \Phi \ll -m^{}_{4} < 0~{\rm eV}$, we instead have 
\begin{eqnarray}\label{eq:wtm1_3}
& & \widetilde{m}^{}_{1} \simeq - \frac{m^{}_{1}+m^{}_{4} +(m^{}_{4} -m^{}_{1})\cos{2\theta^{}_{14}}}{2}+ |g\Phi|   \;, \\
& & \widetilde{m}^{}_{4} \simeq \frac{m^{}_{1}+m^{}_{4}  - (m^{}_{4} -m^{}_{1})\cos{2\theta^{}_{14}}}{2} - \frac{(m^{}_{4} -m^{}_{1})^2 \sin^2{2\theta^{}_{14}}}{4 g \Phi}  \;, \\
& & \tan{2\widetilde{\theta}^{}_{14}} \simeq  \frac{(m^{}_{4} -m^{}_{1}) \sin{2\theta^{}_{14}}}{ g\Phi}\;.
\end{eqnarray}
When $g\Phi$ switches sign, the expressions for $\widetilde{m}^{}_{1}$ and $\widetilde{m}^{}_{4}$ are also swapped, but note that the active neutrino always has dominant overlap with the lighter eigenstate.
The above results are analogous to the analytical observations of standard MSW effect.
In the absence of light sterile neutrino, active neutrinos will develop large effective masses in the early Universe because $\phi$ scales as $(1+z)^{3/2}$ with $z$ being the redshift. This significantly constrains the effect which can be probed in laboratory.
The introduction of a light sterile neutrino is very intriguing in avoiding the CMB limit.
From Eq.~(\ref{eq:wtm1_3}), one can observe that the contribution of large $g\Phi$ to the effective active neutrino mass is negligible, contrary to the case in Eq.~(\ref{eq:wtm1_2}). This can be understood as the fact that the mixing between active and sterile neutrinos is severely suppressed when $g\Phi$ becomes extremely large, whereas $g\Phi$ only acts on the sterile component.
Such a scenario can also be a key to suppress the production of sterile neutrino in the early Universe~\cite{Hagstotz:2020ukm}.

\section{Neutrino oscillations with rapidly oscillating scalar potential}\label{app:C} \noindent
The scenario with slowly oscillating scalar field (compared to the neutrino time of flight) is trivial. Here we explore the scenario with rapid potential alternation during the neutrino time of flight.
Let us focus on the scenario with two neutrino flavors as before, i.e., $\nu^{}_{\rm a}$ and $N^{\rm c}$.
Suppose an active left-handed neutrino component $\nu^{}_{\rm a,L}$ is emitted from the source. We need to figure out how the time-varying neutrino mass alters the prediction of standard neutrino oscillation picture. The analysis present below should also be applicable to time-varying masses of active neutrinos only.

We start by explaining that neutrino-antineutrino oscillation, e.g., $\nu^{}_{\rm a,  L} \to \overline{\nu}^{}_{\rm a, R}$, is always negligible in our scenario.
Since we are interested in light sterile neutrinos here, all mass terms are tiny in comparison to the neutrino beam energy $E^{}_{\nu} \gtrsim 1~{\rm MeV}$. We further assume that the scalar potential is also negligible compared to $E^{}_{\nu}$, such that neutrinos are always ultra-relativistic during the flight. This guarantees that a left-helicity state is always dominantly composed of the left-handed field.
In the mean time, during the propagation the neutrino helicity is a conserved quantum number, because the addition of the time-varying potential $\Phi(t)$ does not affect the spatial translation symmetry (conserving three momentum) and isotropy (conserving angular momentum).
To see it more explicitly, we write down the quantum mechanical Hamiltonian driving the two-component spinors in the basis of $(\nu^{}_{\rm a, L},\nu^{\rm c}_{\rm a, L},N^{\rm c}, N)$:
\begin{eqnarray}
\mathcal{H}(t) =  
\left( \begin{matrix}
\mathrm{i} \bm{\sigma} \cdot \bm{\partial} & m^{}_{\nu} & 0 & m^{}_{\rm D} \\
m^{}_{\nu} & -\mathrm{i} \bm{\sigma} \cdot \bm{\partial} & m^{}_{\rm D} & 0 \\
0 & m^{}_{\rm D} & \mathrm{i} \bm{\sigma} \cdot \bm{\partial} & m^{}_{N}+ g\Phi \\
m^{}_{\rm D} & 0  & m^{}_{N}+ g\Phi & -\mathrm{i} \bm{\sigma} \cdot \bm{\partial} \\
\end{matrix} \right).  \hspace{0.1cm}
\end{eqnarray}
The helicity operator is given by $\bm{\hat{p}} \cdot \bm{\Sigma}$ with $\bm{\hat{p}}$ being the momentum direction vector and $\bm{\Sigma} = \bm{\sigma} \otimes \mathbb{1}^{}_{4 \times 4}$. One can easily check that $[H, \bm{\hat{p}} \cdot \bm{\Sigma}] = 0$ by noticing $[\bm{\hat{p}} \cdot \bm{\sigma} , \bm{\sigma} \cdot \bm{\partial} ] = 0 $.
As a consequence, the neutrino produced from the source, which is dominantly left-helicity, will project mainly into the left-handed field operator at the detector.
The right-handed component is always negligible providing that all the mass  terms and scalar potential are small compared to the neutrino beam energy.
We have also verified the behavior numerically.

Hence, in the following we shall confine the discussion to the flavor conversion between fields with the same handness (say, left handness).
In the flavor basis, the approximated Hamiltonian for a plane wave reads
\begin{eqnarray}
H(t) = p + \frac{M^{}_{\nu} M^{\dagger}_{\nu}(t)}{2p} \;.
\end{eqnarray}
The Hamiltonian relevant for active-sterile neutrino oscillations is reduced to a vacuum term and a time-dependent one, namely $H= H^{}_{0} + H^{}_{\rm I}$ with
\begin{eqnarray}
H^{}_{0} & = &  \frac{1}{2p}\left[\begin{matrix}
m^2_{\nu} + m^2_{\rm D} & m^{}_{\rm D}(m^{}_{\nu}+m^{}_{N} ) \\
m^{}_{\rm D}(m^{}_{\nu}+m^{}_{N} ) &  m^2_{\rm D}+ m^{2}_{N}
\end{matrix} \right] ,  \\ 
H^{}_{\rm I}(t) & = &  \frac{1}{2p}\left[\begin{matrix}
0 & m^{}_{\rm D} \, g \Phi\\
m^{}_{\rm D}\, g\Phi & 2 m^{}_{N} g \Phi+(g \Phi)^2
\end{matrix} \right]. \label{eq:HIt}
\end{eqnarray}
The evolution of neutrino state $| \nu \rangle^{}_{t} = C^{}_{\rm a} | \nu^{}_{\rm a} \rangle + C^{}_{N} | N^{\rm c} \rangle$ is governed by the equation
\begin{eqnarray}
\mathrm{i} \frac{\partial{}}{\partial{t}} (C^{}_{\rm a}, C^{}_{N})^{\rm T} = H(t) (C^{}_{\rm a}, C^{}_{N})^{\rm T} \;.
\end{eqnarray} 
Here, $|C^{}_{\rm a}|^2$ gives the survival probability of $\nu^{}_{\rm a}$. 
When $H^{}_{\rm I}$ is tiny compared to $H^{}_{0}$, one may consider to perturbatively analyze the effect of $g \phi$ by treating $H^{}_{\rm I}$ as a small parameter.
However, we note that neutrino oscillation is an accumulative effect over a macroscopic baseline $L$, and small terms accumulated over $L$ are not always appropriate to be treated as perturbations.
In a general circumstance, exact calculation should be invoked to derive the oscillation probability.

For a rapidly oscillating scalar field with $m^{}_{\phi} \gg \Delta m^2 /(2E)$, neutrinos  feel an averaged potential during propagation. This does not mean that the scalar effect is vanishing. In fact, the average of $H^{}_{\rm I}(t)$ generates a non-vanishing contribution
\begin{eqnarray}
\left\langle H^{}_{\rm I}(t) \right\rangle & = &  \frac{1}{2p}\left[\begin{matrix}
0 & 0 \\
0 & (g \phi)^2/2
\end{matrix} \right],
\end{eqnarray}
which affects neutrino oscillations by modifying the effective mass-squared difference and mixing angle.
The ultimate effective Hamiltonian will be $H^{}_{0} + \left\langle H^{}_{\rm I}(t) \right\rangle$.
The above results should also be valid for decoherent scalar field with random phases, just as the normal matter effect works for an assemble of random microscopic particles.
\bibliographystyle{utcaps_mod}

\bibliography{references}

\providecommand{\href}[2]{#2}\begingroup\raggedright\begin{thebibliography}{100}

\bibitem{Cowan:1956rrn}
C.~L. Cowan, F.~Reines, F.~B. Harrison, H.~W. Kruse, and A.~D. McGuire, ``{\em
  {Detection of the free neutrino: A Confirmation}},''
  \href{http://dx.doi.org/10.1126/science.124.3212.103}{Science {\normalfont
  \bfseries 124} (1956)  103--104}.

\bibitem{Reines:1953pu}
F.~Reines and C.~L. Cowan, ``{\em {Detection of the free neutrino}},''
  \href{http://dx.doi.org/10.1103/PhysRev.92.830}{Phys. Rev. {\normalfont
  \bfseries 92} (1953)  830--831}.

\bibitem{Athar:2021xsd}
M.~S. Athar {\em et al.}, ``{\em {Status and perspectives of neutrino
  physics}},'' \href{http://dx.doi.org/10.1016/j.ppnp.2022.103947}{Prog. Part.
  Nucl. Phys. {\normalfont \bfseries 124} (2022)  103947},
  \href{http://arxiv.org/abs/2111.07586}{{\normalfont \ttfamily
  arXiv:2111.07586}}.

\bibitem{ParticleDataGroup:2020ssz}
{\normalfont \bfseries Particle Data Group}, P.~A. Zyla {\em et al.}, ``{\em
  {Review of Particle Physics}},''
  \href{http://dx.doi.org/10.1093/ptep/ptaa104}{PTEP {\normalfont \bfseries
  2020} (2020) no.~8, 083C01}.

\bibitem{Fermi:1934hr}
E.~Fermi, ``{\em {An attempt of a theory of beta radiation. 1.}},''
\href{http://dx.doi.org/10.1007/BF01351864}{Z. Phys. {\normalfont \bfseries 88}
  (1934)  161--177}.
%%CITATION = ZEPYA,88,161;%%.

\bibitem{Fermi:1934sk}
E.~Fermi\vspace{0mm}, ``{\em {Trends to a Theory of beta Radiation. (In
  Italian)}},''
\href{http://dx.doi.org/10.1007/BF02959820}{Nuovo Cim. {\normalfont \bfseries
  11} (1934)  1--19}.
%%CITATION = NUCIA,11,1;%%.

\bibitem{Perrin:1933}
F.~Perrin, ``{\em {Possibility of emission of neutral particles with zero
  intrinsic mass in beta radioactivity. (In French)}},'' Comptes Rendus
  {\normalfont \bfseries 197} (1933)  1625.

\bibitem{Robertson:1991vn}
R.~G.~H. Robertson, T.~J. Bowles, G.~J. Stephenson, D.~L. Wark, J.~F.
  Wilkerson, and D.~A. Knapp, ``{\em {Limit on anti-electron-neutrino mass from
  observation of the beta decay of molecular tritium}},''
  \href{http://dx.doi.org/10.1103/PhysRevLett.67.957}{Phys. Rev. Lett.
  {\normalfont \bfseries 67} (1991)  957--960}.

\bibitem{Holzschuh:1992np}
E.~Holzschuh, M.~Fritschi, and W.~Kuendig, ``{\em {Measurement of the
  electron-neutrino mass from tritium beta decay}},''
  \href{http://dx.doi.org/10.1016/0370-2693(92)91000-Y}{Phys. Lett. B
  {\normalfont \bfseries 287} (1992)  381--388}.

\bibitem{Kawakami:1991th}
H.~Kawakami {\em et al.}, ``{\em {New upper bound on the electron anti-neutrino
  mass}},'' \href{http://dx.doi.org/10.1016/0370-2693(91)90226-G}{Phys. Lett. B
  {\normalfont \bfseries 256} (1991)  105--111}.

\bibitem{Sun:1993}
H.~Sun {\em et al.} CJNP {\normalfont \bfseries 15} (1993)  261.

\bibitem{Stoeffl:1995wm}
W.~Stoeffl and D.~J. Decman, ``{\em {Anomalous Structure in the Beta Decay of
  Gaseous Molecular Tritium}},''
  \href{http://dx.doi.org/10.1103/PhysRevLett.75.3237}{Phys. Rev. Lett.
  {\normalfont \bfseries 75} (1995)  3237--3240}.

\bibitem{Kraus:2004zw}
C.~Kraus {\em et al.}, ``{\em {Final results from phase II of the Mainz
  neutrino mass search in tritium beta decay}},''
  \href{http://dx.doi.org/10.1140/epjc/s2005-02139-7}{Eur. Phys. J. C
  {\normalfont \bfseries 40} (2005)  447--468},
  \href{http://arxiv.org/abs/hep-ex/0412056}{{\normalfont \ttfamily
  arXiv:hep-ex/0412056}}.

\bibitem{Troitsk:2011cvm}
{\normalfont \bfseries Troitsk}, V.~N. Aseev {\em et al.}, ``{\em {An upper
  limit on electron antineutrino mass from Troitsk experiment}},''
  \href{http://dx.doi.org/10.1103/PhysRevD.84.112003}{Phys. Rev. D {\normalfont
  \bfseries 84} (2011)  112003},
  \href{http://arxiv.org/abs/1108.5034}{{\normalfont \ttfamily
  arXiv:1108.5034}}.

\bibitem{KATRIN:2019yun}
{\normalfont \bfseries KATRIN}, M.~Aker {\em et al.}, ``{\em {Improved Upper
  Limit on the Neutrino Mass from a Direct Kinematic Method by KATRIN}},''
  \href{http://dx.doi.org/10.1103/PhysRevLett.123.221802}{Phys. Rev. Lett.
  {\normalfont \bfseries 123} (2019) no.~22, 221802},
  \href{http://arxiv.org/abs/1909.06048}{{\normalfont \ttfamily
  arXiv:1909.06048}}.

\bibitem{KATRIN:2021fgc}
{\normalfont \bfseries KATRIN}, M.~Aker {\em et al.}, ``{\em {Analysis methods
  for the first KATRIN neutrino-mass measurement}},''
  \href{http://dx.doi.org/10.1103/PhysRevD.104.012005}{Phys. Rev. D
  {\normalfont \bfseries 104} (2021) no.~1, 012005},
  \href{http://arxiv.org/abs/2101.05253}{{\normalfont \ttfamily
  arXiv:2101.05253}}.

\bibitem{KATRIN:2021uub}
{\normalfont \bfseries KATRIN}, M.~Aker {\em et al.}, ``{\em {Direct
  neutrino-mass measurement with sub-electronvolt sensitivity}},''
  \href{http://dx.doi.org/10.1038/s41567-021-01463-1}{Nature Phys. {\normalfont
  \bfseries 18} (2022) no.~2, 160--166},
  \href{http://arxiv.org/abs/2105.08533}{{\normalfont \ttfamily
  arXiv:2105.08533}}.

\bibitem{KATRIN:2022ayy}
{\normalfont \bfseries KATRIN}, M.~Aker {\em et al.}, ``{\em {KATRIN: Status
  and Prospects for the Neutrino Mass and Beyond}},''
  \href{http://arxiv.org/abs/2203.08059}{{\normalfont \ttfamily
  arXiv:2203.08059}}.

\bibitem{Project8:2022wqh}
{\normalfont \bfseries Project 8}, A.~A. Esfahani {\em et al.}, ``{\em {The
  Project 8 Neutrino Mass Experiment}},'' in {\em {2022 Snowmass Summer
  Study}}.
\newblock 3, 2022.
\newblock \href{http://arxiv.org/abs/2203.07349}{{\normalfont \ttfamily
  arXiv:2203.07349}}.

\bibitem{Ma:1998dn}
E.~Ma, ``{\em {Pathways to naturally small neutrino masses}},''
  \href{http://dx.doi.org/10.1103/PhysRevLett.81.1171}{Phys. Rev. Lett.
  {\normalfont \bfseries 81} (1998)  1171--1174},
  \href{http://arxiv.org/abs/hep-ph/9805219}{{\normalfont \ttfamily
  arXiv:hep-ph/9805219}}.

\bibitem{Cai:2017jrq}
Y.~Cai, J.~Herrero-García, M.~A. Schmidt, A.~Vicente, and R.~R. Volkas, ``{\em
  {From the trees to the forest: a review of radiative neutrino mass
  models}},'' \href{http://dx.doi.org/10.3389/fphy.2017.00063}{Front. in Phys.
  {\normalfont \bfseries 5} (2017)  63},
  \href{http://arxiv.org/abs/1706.08524}{{\normalfont \ttfamily
  arXiv:1706.08524}}.

\bibitem{Cai:2017mow}
Y.~Cai, T.~Han, T.~Li, and R.~Ruiz, ``{\em {Lepton Number Violation: Seesaw
  Models and Their Collider Tests}},''
  \href{http://dx.doi.org/10.3389/fphy.2018.00040}{Front. in Phys. {\normalfont
  \bfseries 6} (2018)  40}, \href{http://arxiv.org/abs/1711.02180}{{\normalfont
  \ttfamily arXiv:1711.02180}}.

\bibitem{Fritzsch:1975sr}
H.~Fritzsch, M.~Gell-Mann, and P.~Minkowski, ``{\em {Vector - Like Weak
  Currents and New Elementary Fermions}},''
\href{http://dx.doi.org/10.1016/0370-2693(75)90040-4}{Phys. Lett. {\normalfont
  \bfseries 59B} (1975)  256--260}.
%%CITATION = PHLTA,59B,256;%%.

\bibitem{Cheng:1975gk}
T.~P. Cheng,
``{\em {Lepton Number Nonconservation Effects in the Vector-Like Weak
  Interaction Theory}},''.
%%CITATION = PRINT-75-0803 (MISSOURI);%%.

\bibitem{Fritzsch:1975rz}
H.~Fritzsch and P.~Minkowski, ``{\em {Vector-Like Weak Currents, Massive
  Neutrinos, and Neutrino Beam Oscillations}},''
\href{http://dx.doi.org/10.1016/0370-2693(76)90051-4}{Phys. Lett. {\normalfont
  \bfseries 62B} (1976)  72--76}.
%%CITATION = PHLTA,62B,72;%%.

\bibitem{Minkowski:1977sc}
P.~Minkowski, ``{\em {$\mu \to e\gamma$ at a Rate of One Out of $10^{9}$ Muon
  Decays?}},''
\href{http://dx.doi.org/10.1016/0370-2693(77)90435-X}{Phys. Lett. {\normalfont
  \bfseries 67B} (1977)  421--428}.
%%CITATION = PHLTA,67B,421;%%.

\bibitem{Yanagida:1980xy}
T.~Yanagida, ``{\em {Horizontal Symmetry and Masses of Neutrinos}},''
\href{http://dx.doi.org/10.1143/PTP.64.1103}{Prog. Theor. Phys. {\normalfont
  \bfseries 64} (1980)  1103}.
%%CITATION = PTPKA,64,1103;%%.

\bibitem{GellMann:1980vs}
M.~Gell-Mann, P.~Ramond, and R.~Slansky, ``{\em {Complex Spinors and Unified
  Theories}},'' Conf. Proc. {\normalfont \bfseries C790927} (1979)  315--321,
\href{http://arxiv.org/abs/1306.4669}{{\normalfont \ttfamily arXiv:1306.4669}}.
%%CITATION = ARXIV:1306.4669;%%.

\bibitem{Mohapatra:1979ia}
R.~N. Mohapatra and G.~Senjanovic, ``{\em {Neutrino Mass and Spontaneous Parity
  Nonconservation}},''
  \href{http://dx.doi.org/10.1103/PhysRevLett.44.912}{Phys. Rev. Lett.
  {\normalfont \bfseries 44} (1980)  912}.
[,231(1979)].
%%CITATION = PRLTA,44,912;%%.

\bibitem{Mohapatra:1980yp}
R.~N. Mohapatra and G.~Senjanovic, ``{\em {Neutrino Masses and Mixings in Gauge
  Models with Spontaneous Parity Violation}},''
\href{http://dx.doi.org/10.1103/PhysRevD.23.165}{Phys. Rev. {\normalfont
  \bfseries D23} (1981)  165}.
%%CITATION = PHRVA,D23,165;%%.

\bibitem{Lazarides:1980nt}
G.~Lazarides, Q.~Shafi, and C.~Wetterich, ``{\em {Proton Lifetime and Fermion
  Masses in an SO(10) Model}},''
\href{http://dx.doi.org/10.1016/0550-3213(81)90354-0}{Nucl. Phys. {\normalfont
  \bfseries B181} (1981)  287--300}.
%%CITATION = NUPHA,B181,287;%%.

\bibitem{Konetschny:1977bn}
W.~Konetschny and W.~Kummer, ``{\em {Nonconservation of Total Lepton Number
  with Scalar Bosons}},''
\href{http://dx.doi.org/10.1016/0370-2693(77)90407-5}{Phys. Lett. {\normalfont
  \bfseries 70B} (1977)  433--435}.
%%CITATION = PHLTA,70B,433;%%.

\bibitem{Magg:1980ut}
M.~Magg and C.~Wetterich, ``{\em {Neutrino Mass Problem and Gauge
  Hierarchy}},''
\href{http://dx.doi.org/10.1016/0370-2693(80)90825-4}{Phys. Lett. {\normalfont
  \bfseries 94B} (1980)  61--64}.
%%CITATION = PHLTA,94B,61;%%.

\bibitem{Schechter:1980gr}
J.~Schechter and J.~W.~F. Valle, ``{\em {Neutrino Masses in SU(2) x U(1)
  Theories}},''
\href{http://dx.doi.org/10.1103/PhysRevD.22.2227}{Phys. Rev. {\normalfont
  \bfseries D22} (1980)  2227}.
%%CITATION = PHRVA,D22,2227;%%.

\bibitem{Cheng:1980qt}
T.~P. Cheng and L.-F. Li, ``{\em {Neutrino Masses, Mixings and Oscillations in
  SU(2) x U(1) Models of Electroweak Interactions}},''
\href{http://dx.doi.org/10.1103/PhysRevD.22.2860}{Phys. Rev. {\normalfont
  \bfseries D22} (1980)  2860}.
%%CITATION = PHRVA,D22,2860;%%.

\bibitem{Foot:1988aq}
R.~Foot, H.~Lew, X.~G. He, and G.~C. Joshi, ``{\em {Seesaw Neutrino Masses
  Induced by a Triplet of Leptons}},''
\href{http://dx.doi.org/10.1007/BF01415558}{Z. Phys. {\normalfont \bfseries
  C44} (1989)  441}.
%%CITATION = ZEPYA,C44,441;%%.

\bibitem{Frieman:1995pm}
J.~A. Frieman, C.~T. Hill, A.~Stebbins, and I.~Waga, ``{\em {Cosmology with
  ultralight pseudo Nambu-Goldstone bosons}},''
  \href{http://dx.doi.org/10.1103/PhysRevLett.75.2077}{Phys. Rev. Lett.
  {\normalfont \bfseries 75} (1995)  2077--2080},
  \href{http://arxiv.org/abs/astro-ph/9505060}{{\normalfont \ttfamily
  arXiv:astro-ph/9505060}}.

\bibitem{Fardon:2003eh}
R.~Fardon, A.~E. Nelson, and N.~Weiner, ``{\em {Dark energy from mass varying
  neutrinos}},'' \href{http://dx.doi.org/10.1088/1475-7516/2004/10/005}{JCAP
  {\normalfont \bfseries 10} (2004)  005},
  \href{http://arxiv.org/abs/astro-ph/0309800}{{\normalfont \ttfamily
  arXiv:astro-ph/0309800}}.

\bibitem{Davoudiasl:2018hjw}
H.~Davoudiasl, G.~Mohlabeng, and M.~Sullivan, ``{\em {Galactic Dark Matter
  Population as the Source of Neutrino Masses}},''
  \href{http://dx.doi.org/10.1103/PhysRevD.98.021301}{Phys. Rev. D {\normalfont
  \bfseries 98} (2018) no.~2, 021301},
  \href{http://arxiv.org/abs/1803.00012}{{\normalfont \ttfamily
  arXiv:1803.00012}}.

\bibitem{Ferreira:2020fam}
E.~G.~M. Ferreira, ``{\em {Ultra-light dark matter}},''
  \href{http://dx.doi.org/10.1007/s00159-021-00135-6}{Astron. Astrophys. Rev.
  {\normalfont \bfseries 29} (2021) no.~1, 7},
  \href{http://arxiv.org/abs/2005.03254}{{\normalfont \ttfamily
  arXiv:2005.03254}}.

\bibitem{Urena-Lopez:2019kud}
L.~A. Ure\~na L\'opez, ``{\em {Brief Review on Scalar Field Dark Matter
  Models}},'' \href{http://dx.doi.org/10.3389/fspas.2019.00047}{Front. Astron.
  Space Sci. {\normalfont \bfseries 6} (2019)  47}.

\bibitem{Moore:1994yx}
B.~Moore, ``{\em {Evidence against dissipationless dark matter from
  observations of galaxy haloes}},''
  \href{http://dx.doi.org/10.1038/370629a0}{Nature {\normalfont \bfseries 370}
  (1994)  629}.

\bibitem{Flores:1994gz}
R.~A. Flores and J.~R. Primack, ``{\em {Observational and theoretical
  constraints on singular dark matter halos}},''
  \href{http://dx.doi.org/10.1086/187350}{Astrophys. J. Lett. {\normalfont
  \bfseries 427} (1994)  L1--4},
  \href{http://arxiv.org/abs/astro-ph/9402004}{{\normalfont \ttfamily
  arXiv:astro-ph/9402004}}.

\bibitem{Navarro:1996gj}
J.~F. Navarro, C.~S. Frenk, and S.~D.~M. White, ``{\em {A Universal density
  profile from hierarchical clustering}},''
  \href{http://dx.doi.org/10.1086/304888}{Astrophys. J. {\normalfont \bfseries
  490} (1997)  493--508},
  \href{http://arxiv.org/abs/astro-ph/9611107}{{\normalfont \ttfamily
  arXiv:astro-ph/9611107}}.

\bibitem{Klypin:1999uc}
A.~A. Klypin, A.~V. Kravtsov, O.~Valenzuela, and F.~Prada, ``{\em {Where are
  the missing Galactic satellites?}},''
  \href{http://dx.doi.org/10.1086/307643}{Astrophys. J. {\normalfont \bfseries
  522} (1999)  82--92},
  \href{http://arxiv.org/abs/astro-ph/9901240}{{\normalfont \ttfamily
  arXiv:astro-ph/9901240}}.

\bibitem{Berlin:2016woy}
A.~Berlin, ``{\em {Neutrino Oscillations as a Probe of Light Scalar Dark
  Matter}},'' \href{http://dx.doi.org/10.1103/PhysRevLett.117.231801}{Phys.
  Rev. Lett. {\normalfont \bfseries 117} (2016) no.~23, 231801},
  \href{http://arxiv.org/abs/1608.01307}{{\normalfont \ttfamily
  arXiv:1608.01307}}.

\bibitem{Brdar:2017kbt}
V.~Brdar, J.~Kopp, J.~Liu, P.~Prass, and X.-P. Wang, ``{\em {Fuzzy dark matter
  and nonstandard neutrino interactions}},''
  \href{http://dx.doi.org/10.1103/PhysRevD.97.043001}{Phys. Rev. D {\normalfont
  \bfseries 97} (2018) no.~4, 043001},
  \href{http://arxiv.org/abs/1705.09455}{{\normalfont \ttfamily
  arXiv:1705.09455}}.

\bibitem{Krnjaic:2017zlz}
G.~Krnjaic, P.~A.~N. Machado, and L.~Necib, ``{\em {Distorted neutrino
  oscillations from time varying cosmic fields}},''
  \href{http://dx.doi.org/10.1103/PhysRevD.97.075017}{Phys. Rev. D {\normalfont
  \bfseries 97} (2018) no.~7, 075017},
  \href{http://arxiv.org/abs/1705.06740}{{\normalfont \ttfamily
  arXiv:1705.06740}}.

\bibitem{Zhao:2017wmo}
Y.~Zhao, ``{\em {Cosmology and time dependent parameters induced by a
  misaligned light scalar}},''
  \href{http://dx.doi.org/10.1103/PhysRevD.95.115002}{Phys. Rev. D {\normalfont
  \bfseries 95} (2017) no.~11, 115002},
  \href{http://arxiv.org/abs/1701.02735}{{\normalfont \ttfamily
  arXiv:1701.02735}}.

\bibitem{Liao:2018byh}
J.~Liao, D.~Marfatia, and K.~Whisnant, ``{\em {Light scalar dark matter at
  neutrino oscillation experiments}},''
  \href{http://dx.doi.org/10.1007/JHEP04(2018)136}{JHEP {\normalfont \bfseries
  04} (2018)  136}, \href{http://arxiv.org/abs/1803.01773}{{\normalfont
  \ttfamily arXiv:1803.01773}}.

\bibitem{Capozzi:2018bps}
F.~Capozzi, I.~M. Shoemaker, and L.~Vecchi, ``{\em {Neutrino Oscillations in
  Dark Backgrounds}},''
  \href{http://dx.doi.org/10.1088/1475-7516/2018/07/004}{JCAP {\normalfont
  \bfseries 07} (2018)  004},
  \href{http://arxiv.org/abs/1804.05117}{{\normalfont \ttfamily
  arXiv:1804.05117}}.

\bibitem{Reynoso:2016hjr}
M.~M. Reynoso and O.~A. Sampayo, ``{\em {Propagation of high-energy neutrinos
  in a background of ultralight scalar dark matter}},''
  \href{http://dx.doi.org/10.1016/j.astropartphys.2016.05.004}{Astropart. Phys.
  {\normalfont \bfseries 82} (2016)  10--20},
  \href{http://arxiv.org/abs/1605.09671}{{\normalfont \ttfamily
  arXiv:1605.09671}}.

\bibitem{Huang:2018cwo}
G.-Y. Huang and N.~Nath, ``{\em {Neutrinophilic Axion-Like Dark Matter}},''
  \href{http://dx.doi.org/10.1140/epjc/s10052-018-6391-y}{Eur. Phys. J. C
  {\normalfont \bfseries 78} (2018) no.~11, 922},
  \href{http://arxiv.org/abs/1809.01111}{{\normalfont \ttfamily
  arXiv:1809.01111}}.

\bibitem{Pandey:2018wvh}
S.~Pandey, S.~Karmakar, and S.~Rakshit, ``{\em {Interactions of Astrophysical
  Neutrinos with Dark Matter: A model building perspective}},''
  \href{http://dx.doi.org/10.1007/JHEP01(2019)095}{JHEP {\normalfont \bfseries
  01} (2019)  095}, \href{http://arxiv.org/abs/1810.04203}{{\normalfont
  \ttfamily arXiv:1810.04203}}.

\bibitem{Farzan:2018pnk}
Y.~Farzan and S.~Palomares-Ruiz, ``{\em {Flavor of cosmic neutrinos preserved
  by ultralight dark matter}},''
  \href{http://dx.doi.org/10.1103/PhysRevD.99.051702}{Phys. Rev. D {\normalfont
  \bfseries 99} (2019) no.~5, 051702},
  \href{http://arxiv.org/abs/1810.00892}{{\normalfont \ttfamily
  arXiv:1810.00892}}.

\bibitem{Farzan:2019yvo}
Y.~Farzan, ``{\em {Ultra-light scalar saving the 3 + 1 neutrino scheme from the
  cosmological bounds}},''
  \href{http://dx.doi.org/10.1016/j.physletb.2019.134911}{Phys. Lett. B
  {\normalfont \bfseries 797} (2019)  134911},
  \href{http://arxiv.org/abs/1907.04271}{{\normalfont \ttfamily
  arXiv:1907.04271}}.

\bibitem{Choi:2019ixb}
K.-Y. Choi, J.~Kim, and C.~Rott, ``{\em {Constraining dark matter-neutrino
  interactions with IceCube-170922A}},''
  \href{http://dx.doi.org/10.1103/PhysRevD.99.083018}{Phys. Rev. D {\normalfont
  \bfseries 99} (2019) no.~8, 083018},
  \href{http://arxiv.org/abs/1903.03302}{{\normalfont \ttfamily
  arXiv:1903.03302}}.

\bibitem{Baek:2019wdn}
S.~Baek, ``{\em {Dirac neutrino from the breaking of Peccei-Quinn symmetry}},''
  \href{http://dx.doi.org/10.1016/j.physletb.2020.135415}{Phys. Lett. B
  {\normalfont \bfseries 805} (2020)  135415},
  \href{http://arxiv.org/abs/1911.04210}{{\normalfont \ttfamily
  arXiv:1911.04210}}.

\bibitem{Ge:2019tdi}
S.-F. Ge and H.~Murayama, ``{\em {Apparent CPT Violation in Neutrino
  Oscillation from Dark Non-Standard Interactions}},''
  \href{http://arxiv.org/abs/1904.02518}{{\normalfont \ttfamily
  arXiv:1904.02518}}.

\bibitem{Choi:2019zxy}
K.-Y. Choi, E.~J. Chun, and J.~Kim, ``{\em {Neutrino Oscillations in Dark
  Matter}},'' \href{http://dx.doi.org/10.1016/j.dark.2020.100606}{Phys. Dark
  Univ. {\normalfont \bfseries 30} (2020)  100606},
  \href{http://arxiv.org/abs/1909.10478}{{\normalfont \ttfamily
  arXiv:1909.10478}}.

\bibitem{Choi:2020ydp}
K.-Y. Choi, E.~J. Chun, and J.~Kim, ``{\em {Dispersion of neutrinos in a
  medium}},'' \href{http://arxiv.org/abs/2012.09474}{{\normalfont \ttfamily
  arXiv:2012.09474}}.

\bibitem{Dev:2020kgz}
A.~Dev, P.~A.~N. Machado, and P.~Mart\'\i{}nez-Mirav\'e, ``{\em {Signatures of
  ultralight dark matter in neutrino oscillation experiments}},''
  \href{http://dx.doi.org/10.1007/JHEP01(2021)094}{JHEP {\normalfont \bfseries
  01} (2021)  094}, \href{http://arxiv.org/abs/2007.03590}{{\normalfont
  \ttfamily arXiv:2007.03590}}.

\bibitem{Baek:2020ovw}
S.~Baek, ``{\em {A connection between flavour anomaly, neutrino mass, and
  axion}},'' \href{http://dx.doi.org/10.1007/JHEP10(2020)111}{JHEP {\normalfont
  \bfseries 10} (2020)  111},
  \href{http://arxiv.org/abs/2006.02050}{{\normalfont \ttfamily
  arXiv:2006.02050}}.

\bibitem{Losada:2021bxx}
M.~Losada, Y.~Nir, G.~Perez, and Y.~Shpilman, ``{\em {Probing scalar dark
  matter oscillations with neutrino oscillations}},''
  \href{http://arxiv.org/abs/2107.10865}{{\normalfont \ttfamily
  arXiv:2107.10865}}.

\bibitem{Smirnov:2021zgn}
A.~Y. Smirnov and V.~B. Valera, ``{\em {Resonance refraction and neutrino
  oscillations}},'' \href{http://arxiv.org/abs/2106.13829}{{\normalfont
  \ttfamily arXiv:2106.13829}}.

\bibitem{Alonso-Alvarez:2021pgy}
G.~Alonso-\'Alvarez and J.~M. Cline, ``{\em {Sterile neutrino dark matter
  catalyzed by a very light dark photon}},''
  \href{http://arxiv.org/abs/2107.07524}{{\normalfont \ttfamily
  arXiv:2107.07524}}.

\bibitem{Huang:2021zzz}
G.-y. Huang and W.~Rodejohann, ``{\em {Tritium beta decay with modified
  neutrino dispersion relations: KATRIN in the dark sea}},''
  \href{http://arxiv.org/abs/2110.03718}{{\normalfont \ttfamily
  arXiv:2110.03718}}.

\bibitem{Huang:2021kam}
G.-y. Huang and N.~Nath, ``{\em {Neutrino meets ultralight dark matter: $0
  \nu\beta\beta$ decay and cosmology}},''
  \href{http://arxiv.org/abs/2111.08732}{{\normalfont \ttfamily
  arXiv:2111.08732}}.

\bibitem{Chun:2021ief}
E.~J. Chun, ``{\em {Neutrino Transition in Dark Matter}},''
  \href{http://arxiv.org/abs/2112.05057}{{\normalfont \ttfamily
  arXiv:2112.05057}}.

\bibitem{Reynoso:2022vrn}
M.~M. Reynoso, O.~A. Sampayo, and A.~M. Carulli, ``{\em {Neutrino interactions
  with ultralight axion-like dark matter}},''
  \href{http://dx.doi.org/10.1140/epjc/s10052-022-10228-w}{Eur. Phys. J. C
  {\normalfont \bfseries 82} (2022) no.~3, 274},
  \href{http://arxiv.org/abs/2203.11642}{{\normalfont \ttfamily
  arXiv:2203.11642}}.

\bibitem{Dev:2022bae}
A.~Dev, G.~Krnjaic, P.~Machado, and H.~Ramani, ``{\em {Constraining Feeble
  Neutrino Interactions with Ultralight Dark Matter}},''
  \href{http://arxiv.org/abs/2205.06821}{{\normalfont \ttfamily
  arXiv:2205.06821}}.

\bibitem{Anisimov:2006hv}
A.~Anisimov, \href{http://dx.doi.org/10.1142/9789812770288_0058}{``{\em
  {Majorana Dark Matter}},''} in {\em {6th International Workshop on the
  Identification of Dark Matter}}, pp.~439--449.
\newblock 12, 2006.
\newblock \href{http://arxiv.org/abs/hep-ph/0612024}{{\normalfont \ttfamily
  arXiv:hep-ph/0612024}}.

\bibitem{Anisimov:2008gg}
A.~Anisimov and P.~Di~Bari, ``{\em {Cold Dark Matter from heavy Right-Handed
  neutrino mixing}},''
  \href{http://dx.doi.org/10.1103/PhysRevD.80.073017}{Phys. Rev. D {\normalfont
  \bfseries 80} (2009)  073017},
  \href{http://arxiv.org/abs/0812.5085}{{\normalfont \ttfamily
  arXiv:0812.5085}}.

\bibitem{Chikashige:1980qk}
Y.~Chikashige, R.~N. Mohapatra, and R.~D. Peccei, ``{\em {Spontaneously Broken
  Lepton Number and Cosmological Constraints on the Neutrino Mass Spectrum}},''
  \href{http://dx.doi.org/10.1103/PhysRevLett.45.1926}{Phys. Rev. Lett.
  {\normalfont \bfseries 45} (1980)  1926}.

\bibitem{Chikashige:1980ui}
Y.~Chikashige, R.~N. Mohapatra, and R.~D. Peccei, ``{\em {Are There Real
  Goldstone Bosons Associated with Broken Lepton Number?}},''
  \href{http://dx.doi.org/10.1016/0370-2693(81)90011-3}{Phys. Lett. B
  {\normalfont \bfseries 98} (1981)  265--268}.

\bibitem{Gelmini:1980re}
G.~B. Gelmini and M.~Roncadelli, ``{\em {Left-Handed Neutrino Mass Scale and
  Spontaneously Broken Lepton Number}},''
  \href{http://dx.doi.org/10.1016/0370-2693(81)90559-1}{Phys. Lett. B
  {\normalfont \bfseries 99} (1981)  411--415}.

\bibitem{Choi:1991aa}
K.~Choi and A.~Santamaria, ``{\em {17-KeV neutrino in a singlet - triplet
  majoron model}},''
  \href{http://dx.doi.org/10.1016/0370-2693(91)90900-B}{Phys. Lett. B
  {\normalfont \bfseries 267} (1991)  504--508}.

\bibitem{Acker:1992eh}
A.~Acker, A.~Joshipura, and S.~Pakvasa, ``{\em {A Neutrino decay model, solar
  anti-neutrinos and atmospheric neutrinos}},''
  \href{http://dx.doi.org/10.1016/0370-2693(92)91520-J}{Phys. Lett. B
  {\normalfont \bfseries 285} (1992)  371--375}.

\bibitem{Georgi:1981pg}
H.~M. Georgi, S.~L. Glashow, and S.~Nussinov, ``{\em {Unconventional Model of
  Neutrino Masses}},''
  \href{http://dx.doi.org/10.1016/0550-3213(81)90336-9}{Nucl. Phys. B
  {\normalfont \bfseries 193} (1981)  297--316}.

\bibitem{Schechter:1981cv}
J.~Schechter and J.~W.~F. Valle, ``{\em {Neutrino Decay and Spontaneous
  Violation of Lepton Number}},''
  \href{http://dx.doi.org/10.1103/PhysRevD.25.774}{Phys. Rev. D {\normalfont
  \bfseries 25} (1982)  774}.

\bibitem{Brookfield:2005bz}
A.~W. Brookfield, C.~van~de Bruck, D.~F. Mota, and D.~Tocchini-Valentini,
  ``{\em {Cosmology of mass-varying neutrinos driven by quintessence: theory
  and observations}},''
  \href{http://dx.doi.org/10.1103/PhysRevD.73.083515}{Phys. Rev. D {\normalfont
  \bfseries 73} (2006)  083515},
  \href{http://arxiv.org/abs/astro-ph/0512367}{{\normalfont \ttfamily
  arXiv:astro-ph/0512367}}. [Erratum: Phys.Rev.D 76, 049901 (2007)].

\bibitem{Lorenz:2018fzb}
C.~S. Lorenz, L.~Funcke, E.~Calabrese, and S.~Hannestad, ``{\em {Time-varying
  neutrino mass from a supercooled phase transition: current cosmological
  constraints and impact on the $\Omega_m$-$\sigma_8$ plane}},''
  \href{http://dx.doi.org/10.1103/PhysRevD.99.023501}{Phys. Rev. D {\normalfont
  \bfseries 99} (2019) no.~2, 023501},
  \href{http://arxiv.org/abs/1811.01991}{{\normalfont \ttfamily
  arXiv:1811.01991}}.

\bibitem{Dvali:2016uhn}
G.~Dvali and L.~Funcke, ``{\em {Small neutrino masses from gravitational
  \ensuremath{\theta}-term}},''
  \href{http://dx.doi.org/10.1103/PhysRevD.93.113002}{Phys. Rev. D {\normalfont
  \bfseries 93} (2016) no.~11, 113002},
  \href{http://arxiv.org/abs/1602.03191}{{\normalfont \ttfamily
  arXiv:1602.03191}}.

\bibitem{Lorenz:2021alz}
C.~S. Lorenz, L.~Funcke, M.~L\"offler, and E.~Calabrese, ``{\em {Reconstruction
  of the neutrino mass as a function of redshift}},''
  \href{http://dx.doi.org/10.1103/PhysRevD.104.123518}{Phys. Rev. D
  {\normalfont \bfseries 104} (2021) no.~12, 123518},
  \href{http://arxiv.org/abs/2102.13618}{{\normalfont \ttfamily
  arXiv:2102.13618}}.

\bibitem{deGouvea:2022dtw}
A.~de~Gouv\^ea, I.~Martinez-Soler, Y.~F. Perez-Gonzalez, and M.~Sen, ``{\em
  {The diffuse supernova neutrino background as a probe of late-time neutrino
  mass generation}},'' \href{http://arxiv.org/abs/2205.01102}{{\normalfont
  \ttfamily arXiv:2205.01102}}.

\bibitem{Gariazzo:2015rra}
S.~Gariazzo, C.~Giunti, M.~Laveder, Y.~F. Li, and E.~M. Zavanin, ``{\em {Light
  sterile neutrinos}},''
  \href{http://dx.doi.org/10.1088/0954-3899/43/3/033001}{J. Phys. G
  {\normalfont \bfseries 43} (2016)  033001},
  \href{http://arxiv.org/abs/1507.08204}{{\normalfont \ttfamily
  arXiv:1507.08204}}.

\bibitem{Giunti:2019aiy}
C.~Giunti and T.~Lasserre, ``{\em {eV-scale Sterile Neutrinos}},''
  \href{http://dx.doi.org/10.1146/annurev-nucl-101918-023755}{Ann. Rev. Nucl.
  Part. Sci. {\normalfont \bfseries 69} (2019)  163--190},
  \href{http://arxiv.org/abs/1901.08330}{{\normalfont \ttfamily
  arXiv:1901.08330}}.

\bibitem{Diaz:2019fwt}
A.~Diaz, C.~A. Arg\"uelles, G.~H. Collin, J.~M. Conrad, and M.~H. Shaevitz,
  ``{\em {Where Are We With Light Sterile Neutrinos?}},''
  \href{http://dx.doi.org/10.1016/j.physrep.2020.08.005}{Phys. Rept.
  {\normalfont \bfseries 884} (2020)  1--59},
  \href{http://arxiv.org/abs/1906.00045}{{\normalfont \ttfamily
  arXiv:1906.00045}}.

\bibitem{Boser:2019rta}
S.~B\"oser, C.~Buck, C.~Giunti, J.~Lesgourgues, L.~Ludhova, S.~Mertens,
  A.~Schukraft, and M.~Wurm, ``{\em {Status of Light Sterile Neutrino
  Searches}},'' \href{http://dx.doi.org/10.1016/j.ppnp.2019.103736}{Prog. Part.
  Nucl. Phys. {\normalfont \bfseries 111} (2020)  103736},
  \href{http://arxiv.org/abs/1906.01739}{{\normalfont \ttfamily
  arXiv:1906.01739}}.

\bibitem{Gariazzo:2021wsx}
S.~Gariazzo, ``{\em {Light Sterile Neutrinos}},''
  \href{http://dx.doi.org/10.1088/1742-6596/2156/1/012003}{J. Phys. Conf. Ser.
  {\normalfont \bfseries 2156} (2021) no.~1, 012003},
  \href{http://arxiv.org/abs/2110.09876}{{\normalfont \ttfamily
  arXiv:2110.09876}}.

\bibitem{Archidiacono:2022ich}
M.~Archidiacono and S.~Gariazzo, ``{\em {Two Sides of the Same Coin: Sterile
  Neutrinos and Dark Radiation, Status and Perspectives}},''
  \href{http://dx.doi.org/10.3390/universe8030175}{Universe {\normalfont
  \bfseries 8} (2022) no.~3, 175},
  \href{http://arxiv.org/abs/2201.10319}{{\normalfont \ttfamily
  arXiv:2201.10319}}.

\bibitem{Barbieri:1990vx}
R.~Barbieri and A.~Dolgov, ``{\em {Neutrino oscillations in the early
  universe}},'' \href{http://dx.doi.org/10.1016/0550-3213(91)90396-F}{Nucl.
  Phys. B {\normalfont \bfseries 349} (1991)  743--753}.

\bibitem{Enqvist:1990ad}
K.~Enqvist, K.~Kainulainen, and J.~Maalampi, ``{\em {Refraction and
  Oscillations of Neutrinos in the Early Universe}},''
  \href{http://dx.doi.org/10.1016/0550-3213(91)90397-G}{Nucl. Phys. B
  {\normalfont \bfseries 349} (1991)  754--790}.

\bibitem{McKellar:1992ja}
B.~H.~J. McKellar and M.~J. Thomson, ``{\em {Oscillating doublet neutrinos in
  the early universe}},''
  \href{http://dx.doi.org/10.1103/PhysRevD.49.2710}{Phys. Rev. D {\normalfont
  \bfseries 49} (1994)  2710--2728}.

\bibitem{Sigl:1993ctk}
G.~Sigl and G.~Raffelt, ``{\em {General kinetic description of relativistic
  mixed neutrinos}},''
  \href{http://dx.doi.org/10.1016/0550-3213(93)90175-O}{Nucl. Phys. B
  {\normalfont \bfseries 406} (1993)  423--451}.

\bibitem{Hannestad:2012ky}
S.~Hannestad, I.~Tamborra, and T.~Tram, ``{\em {Thermalisation of light sterile
  neutrinos in the early universe}},''
  \href{http://dx.doi.org/10.1088/1475-7516/2012/07/025}{JCAP {\normalfont
  \bfseries 07} (2012)  025},
  \href{http://arxiv.org/abs/1204.5861}{{\normalfont \ttfamily
  arXiv:1204.5861}}.

\bibitem{Gariazzo:2019gyi}
S.~Gariazzo, P.~F. de~Salas, and S.~Pastor, ``{\em {Thermalisation of sterile
  neutrinos in the early Universe in the 3+1 scheme with full mixing
  matrix}},'' \href{http://dx.doi.org/10.1088/1475-7516/2019/07/014}{JCAP
  {\normalfont \bfseries 07} (2019)  014},
  \href{http://arxiv.org/abs/1905.11290}{{\normalfont \ttfamily
  arXiv:1905.11290}}.

\bibitem{Davoudiasl:2019nlo}
H.~Davoudiasl and P.~B. Denton, ``{\em {Ultralight Boson Dark Matter and Event
  Horizon Telescope Observations of M87*}},''
  \href{http://dx.doi.org/10.1103/PhysRevLett.123.021102}{Phys. Rev. Lett.
  {\normalfont \bfseries 123} (2019) no.~2, 021102},
  \href{http://arxiv.org/abs/1904.09242}{{\normalfont \ttfamily
  arXiv:1904.09242}}.

\bibitem{Brito:2015oca}
R.~Brito, V.~Cardoso, and P.~Pani, ``{\em {Superradiance}: {New Frontiers in
  Black Hole Physics}},''
  \href{http://dx.doi.org/10.1007/978-3-319-19000-6}{Lect. Notes Phys.
  {\normalfont \bfseries 906} (2015)  pp.1--237},
  \href{http://arxiv.org/abs/1501.06570}{{\normalfont \ttfamily
  arXiv:1501.06570}}.

\bibitem{Barbieri:1989ti}
R.~Barbieri and A.~Dolgov, ``{\em {Bounds on Sterile-neutrinos from
  Nucleosynthesis}},''
  \href{http://dx.doi.org/10.1016/0370-2693(90)91203-N}{Phys. Lett. B
  {\normalfont \bfseries 237} (1990)  440--445}.

\bibitem{Dolgov:2003sg}
A.~D. Dolgov and F.~L. Villante, ``{\em {BBN bounds on active sterile neutrino
  mixing}},'' \href{http://dx.doi.org/10.1016/j.nuclphysb.2003.11.031}{Nucl.
  Phys. B {\normalfont \bfseries 679} (2004)  261--298},
  \href{http://arxiv.org/abs/hep-ph/0308083}{{\normalfont \ttfamily
  arXiv:hep-ph/0308083}}.

\bibitem{Dasgupta:2013zpn}
B.~Dasgupta and J.~Kopp, ``{\em {Cosmologically Safe eV-Scale Sterile Neutrinos
  and Improved Dark Matter Structure}},''
  \href{http://dx.doi.org/10.1103/PhysRevLett.112.031803}{Phys. Rev. Lett.
  {\normalfont \bfseries 112} (2014) no.~3, 031803},
  \href{http://arxiv.org/abs/1310.6337}{{\normalfont \ttfamily
  arXiv:1310.6337}}.

\bibitem{Hannestad:2013ana}
S.~Hannestad, R.~S. Hansen, and T.~Tram, ``{\em {How Self-Interactions can
  Reconcile Sterile Neutrinos with Cosmology}},''
  \href{http://dx.doi.org/10.1103/PhysRevLett.112.031802}{Phys. Rev. Lett.
  {\normalfont \bfseries 112} (2014) no.~3, 031802},
  \href{http://arxiv.org/abs/1310.5926}{{\normalfont \ttfamily
  arXiv:1310.5926}}.

\bibitem{Mirizzi:2014ama}
A.~Mirizzi, G.~Mangano, O.~Pisanti, and N.~Saviano, ``{\em {Collisional
  production of sterile neutrinos via secret interactions and cosmological
  implications}},'' \href{http://dx.doi.org/10.1103/PhysRevD.91.025019}{Phys.
  Rev. D {\normalfont \bfseries 91} (2015) no.~2, 025019},
  \href{http://arxiv.org/abs/1410.1385}{{\normalfont \ttfamily
  arXiv:1410.1385}}.

\bibitem{Forastieri:2017oma}
F.~Forastieri, M.~Lattanzi, G.~Mangano, A.~Mirizzi, P.~Natoli, and N.~Saviano,
  ``{\em {Cosmic microwave background constraints on secret interactions among
  sterile neutrinos}},''
  \href{http://dx.doi.org/10.1088/1475-7516/2017/07/038}{JCAP {\normalfont
  \bfseries 07} (2017)  038},
  \href{http://arxiv.org/abs/1704.00626}{{\normalfont \ttfamily
  arXiv:1704.00626}}.

\bibitem{Archidiacono:2020yey}
M.~Archidiacono, S.~Gariazzo, C.~Giunti, S.~Hannestad, and T.~Tram, ``{\em
  {Sterile neutrino self-interactions: $H_0$ tension and short-baseline
  anomalies}},'' \href{http://dx.doi.org/10.1088/1475-7516/2020/12/029}{JCAP
  {\normalfont \bfseries 12} (2020)  029},
  \href{http://arxiv.org/abs/2006.12885}{{\normalfont \ttfamily
  arXiv:2006.12885}}.

\bibitem{Dodelson:1993je}
S.~Dodelson and L.~M. Widrow, ``{\em {Sterile-neutrinos as dark matter}},''
  \href{http://dx.doi.org/10.1103/PhysRevLett.72.17}{Phys. Rev. Lett.
  {\normalfont \bfseries 72} (1994)  17--20},
\href{http://arxiv.org/abs/hep-ph/9303287}{{\normalfont \ttfamily
  arXiv:hep-ph/9303287}}.
%%CITATION = HEP-PH/9303287;%%.

\bibitem{Jacques:2013xr}
T.~D. Jacques, L.~M. Krauss, and C.~Lunardini, ``{\em {Additional Light Sterile
  Neutrinos and Cosmology}},''
  \href{http://dx.doi.org/10.1103/PhysRevD.87.083515}{Phys. Rev. D {\normalfont
  \bfseries 87} (2013) no.~8, 083515},
  \href{http://arxiv.org/abs/1301.3119}{{\normalfont \ttfamily
  arXiv:1301.3119}}. [Erratum: Phys.Rev.D 88, 109901 (2013)].

\bibitem{Dessert:2018qih}
C.~Dessert, N.~L. Rodd, and B.~R. Safdi, ``{\em {The dark matter interpretation
  of the 3.5-keV line is inconsistent with blank-sky observations}},''
  \href{http://dx.doi.org/10.1126/science.aaw3772}{Science {\normalfont
  \bfseries 367} (2020) no.~6485, 1465--1467},
  \href{http://arxiv.org/abs/1812.06976}{{\normalfont \ttfamily
  arXiv:1812.06976}}.

\bibitem{Benso:2019jog}
C.~Benso, V.~Brdar, M.~Lindner, and W.~Rodejohann, ``{\em {Prospects for
  Finding Sterile Neutrino Dark Matter at KATRIN}},''
  \href{http://dx.doi.org/10.1103/PhysRevD.100.115035}{Phys. Rev. D
  {\normalfont \bfseries 100} (2019) no.~11, 115035},
  \href{http://arxiv.org/abs/1911.00328}{{\normalfont \ttfamily
  arXiv:1911.00328}}.

\bibitem{Berlin:2016bdv}
A.~Berlin and D.~Hooper, ``{\em {Axion-Assisted Production of Sterile Neutrino
  Dark Matter}},'' \href{http://dx.doi.org/10.1103/PhysRevD.95.075017}{Phys.
  Rev. D {\normalfont \bfseries 95} (2017) no.~7, 075017},
  \href{http://arxiv.org/abs/1610.03849}{{\normalfont \ttfamily
  arXiv:1610.03849}}.

\bibitem{DeGouvea:2019wpf}
A.~De~Gouv\^ea, M.~Sen, W.~Tangarife, and Y.~Zhang, ``{\em {Dodelson-Widrow
  Mechanism in the Presence of Self-Interacting Neutrinos}},''
  \href{http://dx.doi.org/10.1103/PhysRevLett.124.081802}{Phys. Rev. Lett.
  {\normalfont \bfseries 124} (2020) no.~8, 081802},
  \href{http://arxiv.org/abs/1910.04901}{{\normalfont \ttfamily
  arXiv:1910.04901}}.

\bibitem{Planck:2018vyg}
{\normalfont \bfseries Planck}, N.~Aghanim {\em et al.}, ``{\em {Planck 2018
  results. VI. Cosmological parameters}},''
  \href{http://dx.doi.org/10.1051/0004-6361/201833910}{Astron. Astrophys.
  {\normalfont \bfseries 641} (2020)  A6},
  \href{http://arxiv.org/abs/1807.06209}{{\normalfont \ttfamily
  arXiv:1807.06209}}. [Erratum: Astron.Astrophys. 652, C4 (2021)].

\bibitem{Palanque-Delabrouille:2019iyz}
N.~Palanque-Delabrouille, C.~Y\`eche, N.~Sch\"oneberg, J.~Lesgourgues,
  M.~Walther, S.~Chabanier, and E.~Armengaud, ``{\em {Hints, neutrino bounds
  and WDM constraints from SDSS DR14 Lyman-$\alpha$ and Planck full-survey
  data}},'' \href{http://dx.doi.org/10.1088/1475-7516/2020/04/038}{JCAP
  {\normalfont \bfseries 04} (2020)  038},
  \href{http://arxiv.org/abs/1911.09073}{{\normalfont \ttfamily
  arXiv:1911.09073}}.

\bibitem{DiValentino:2021hoh}
E.~Di~Valentino, S.~Gariazzo, and O.~Mena, ``{\em {Most constraining
  cosmological neutrino mass bounds}},''
  \href{http://dx.doi.org/10.1103/PhysRevD.104.083504}{Phys. Rev. D
  {\normalfont \bfseries 104} (2021) no.~8, 083504},
  \href{http://arxiv.org/abs/2106.15267}{{\normalfont \ttfamily
  arXiv:2106.15267}}.

\bibitem{ATLAS:2015gtp}
{\normalfont \bfseries ATLAS}, G.~Aad {\em et al.}, ``{\em {Search for heavy
  Majorana neutrinos with the ATLAS detector in pp collisions at $ \sqrt{s}=8 $
  TeV}},'' \href{http://dx.doi.org/10.1007/JHEP07(2015)162}{JHEP {\normalfont
  \bfseries 07} (2015)  162},
  \href{http://arxiv.org/abs/1506.06020}{{\normalfont \ttfamily
  arXiv:1506.06020}}.

\bibitem{CMS:2016aro}
{\normalfont \bfseries CMS}, V.~Khachatryan {\em et al.}, ``{\em {Search for
  heavy Majorana neutrinos in $e^{\pm} e^{\pm}$ + jets and $e^{\pm} \mu^{\pm}$
  + jets events in proton-proton collisions at $ \sqrt{s}=8 $ TeV}},''
  \href{http://dx.doi.org/10.1007/JHEP04(2016)169}{JHEP {\normalfont \bfseries
  04} (2016)  169}, \href{http://arxiv.org/abs/1603.02248}{{\normalfont
  \ttfamily arXiv:1603.02248}}.

\bibitem{CMS:2018iaf}
{\normalfont \bfseries CMS}, A.~M. Sirunyan {\em et al.}, ``{\em {Search for
  heavy neutral leptons in events with three charged leptons in proton-proton
  collisions at $\sqrt{s} =$ 13 TeV}},''
  \href{http://dx.doi.org/10.1103/PhysRevLett.120.221801}{Phys. Rev. Lett.
  {\normalfont \bfseries 120} (2018) no.~22, 221801},
  \href{http://arxiv.org/abs/1802.02965}{{\normalfont \ttfamily
  arXiv:1802.02965}}.

\bibitem{Kraus:2012he}
C.~Kraus, A.~Singer, K.~Valerius, and C.~Weinheimer, ``{\em {Limit on sterile
  neutrino contribution from the Mainz Neutrino Mass Experiment}},''
  \href{http://dx.doi.org/10.1140/epjc/s10052-013-2323-z}{Eur. Phys. J. C
  {\normalfont \bfseries 73} (2013) no.~2, 2323},
  \href{http://arxiv.org/abs/1210.4194}{{\normalfont \ttfamily
  arXiv:1210.4194}}.

\bibitem{Belesev:2012hx}
A.~I. Belesev, A.~I. Berlev, E.~V. Geraskin, A.~A. Golubev, N.~A. Likhovid,
  A.~A. Nozik, V.~S. Pantuev, V.~I. Parfenov, and A.~K. Skasyrskaya, ``{\em {An
  upper limit on additional neutrino mass eigenstate in 2 to 100 eV region from
  'Troitsk nu-mass' data}},''
  \href{http://dx.doi.org/10.1134/S0021364013020033}{JETP Lett. {\normalfont
  \bfseries 97} (2013)  67--69},
  \href{http://arxiv.org/abs/1211.7193}{{\normalfont \ttfamily
  arXiv:1211.7193}}.

\bibitem{Belesev:2013cba}
A.~I. Belesev, A.~I. Berlev, E.~V. Geraskin, A.~A. Golubev, N.~A. Likhovid,
  A.~A. Nozik, V.~S. Pantuev, V.~I. Parfenov, and A.~K. Skasyrskaya, ``{\em
  {The search for an additional neutrino mass eigenstate in the
  2\textendash{}100 eV region from \textquoteleft{}Troitsk
  nu-mass\textquoteright{} data: a detailed analysis}},''
  \href{http://dx.doi.org/10.1088/0954-3899/41/1/015001}{J. Phys. G
  {\normalfont \bfseries 41} (2014)  015001},
  \href{http://arxiv.org/abs/1307.5687}{{\normalfont \ttfamily
  arXiv:1307.5687}}.

\bibitem{Alvey:2021xmq}
J.~Alvey, M.~Escudero, N.~Sabti, and T.~Schwetz, ``{\em {Cosmic neutrino
  background detection in large-neutrino-mass cosmologies}},''
  \href{http://dx.doi.org/10.1103/PhysRevD.105.063501}{Phys. Rev. D
  {\normalfont \bfseries 105} (2022) no.~6, 063501},
  \href{http://arxiv.org/abs/2111.14870}{{\normalfont \ttfamily
  arXiv:2111.14870}}.

\bibitem{KATRIN:2020dpx}
{\normalfont \bfseries KATRIN}, M.~Aker {\em et al.}, ``{\em {Bound on 3+1
  Active-Sterile Neutrino Mixing from the First Four-Week Science Run of
  KATRIN}},'' \href{http://dx.doi.org/10.1103/PhysRevLett.126.091803}{Phys.
  Rev. Lett. {\normalfont \bfseries 126} (2021) no.~9, 091803},
  \href{http://arxiv.org/abs/2011.05087}{{\normalfont \ttfamily
  arXiv:2011.05087}}.

\bibitem{Giunti:2019fcj}
C.~Giunti, Y.~F. Li, and Y.~Y. Zhang, ``{\em {KATRIN bound on 3+1
  active-sterile neutrino mixing and the reactor antineutrino anomaly}},''
  \href{http://dx.doi.org/10.1007/JHEP05(2020)061}{JHEP {\normalfont \bfseries
  05} (2020)  061}, \href{http://arxiv.org/abs/1912.12956}{{\normalfont
  \ttfamily arXiv:1912.12956}}.

\bibitem{KATRIN:2022ith}
{\normalfont \bfseries KATRIN}, M.~Aker {\em et al.}, ``{\em {Improved eV-scale
  sterile-neutrino constraints from the second KATRIN measurement campaign}},''
  \href{http://dx.doi.org/10.1103/PhysRevD.105.072004}{Phys. Rev. D
  {\normalfont \bfseries 105} (2022) no.~7, 072004},
  \href{http://arxiv.org/abs/2201.11593}{{\normalfont \ttfamily
  arXiv:2201.11593}}.

\bibitem{Kaether:2010ag}
F.~Kaether, W.~Hampel, G.~Heusser, J.~Kiko, and T.~Kirsten, ``{\em {Reanalysis
  of the GALLEX solar neutrino flux and source experiments}},''
  \href{http://dx.doi.org/10.1016/j.physletb.2010.01.030}{Phys. Lett. B
  {\normalfont \bfseries 685} (2010)  47--54},
  \href{http://arxiv.org/abs/1001.2731}{{\normalfont \ttfamily
  arXiv:1001.2731}}.

\bibitem{SAGE:2009eeu}
{\normalfont \bfseries SAGE}, J.~N. Abdurashitov {\em et al.}, ``{\em
  {Measurement of the solar neutrino capture rate with gallium metal. III:
  Results for the 2002--2007 data-taking period}},''
  \href{http://dx.doi.org/10.1103/PhysRevC.80.015807}{Phys. Rev. C {\normalfont
  \bfseries 80} (2009)  015807},
  \href{http://arxiv.org/abs/0901.2200}{{\normalfont \ttfamily
  arXiv:0901.2200}}.

\bibitem{Barinov:2021asz}
V.~V. Barinov {\em et al.}, ``{\em {Results from the Baksan Experiment on
  Sterile Transitions (BEST)}},''
  \href{http://arxiv.org/abs/2109.11482}{{\normalfont \ttfamily
  arXiv:2109.11482}}.

\bibitem{Barinov:2022wfh}
V.~V. Barinov {\em et al.}, ``{\em {A Search for Electron Neutrino Transitions
  to Sterile States in the BEST Experiment}},''
  \href{http://arxiv.org/abs/2201.07364}{{\normalfont \ttfamily
  arXiv:2201.07364}}.

\bibitem{Kopp:2013vaa}
J.~Kopp, P.~A.~N. Machado, M.~Maltoni, and T.~Schwetz, ``{\em {Sterile Neutrino
  Oscillations: The Global Picture}},''
  \href{http://dx.doi.org/10.1007/JHEP05(2013)050}{JHEP {\normalfont \bfseries
  05} (2013)  050}, \href{http://arxiv.org/abs/1303.3011}{{\normalfont
  \ttfamily arXiv:1303.3011}}.

\bibitem{Dentler:2018sju}
M.~Dentler, A.~Hern\'andez-Cabezudo, J.~Kopp, P.~A.~N. Machado, M.~Maltoni,
  I.~Martinez-Soler, and T.~Schwetz, ``{\em {Updated Global Analysis of
  Neutrino Oscillations in the Presence of eV-Scale Sterile Neutrinos}},''
  \href{http://dx.doi.org/10.1007/JHEP08(2018)010}{JHEP {\normalfont \bfseries
  08} (2018)  010}, \href{http://arxiv.org/abs/1803.10661}{{\normalfont
  \ttfamily arXiv:1803.10661}}.

\bibitem{DayaBay:2018fsh}
{\normalfont \bfseries Daya Bay}, D.~Adey {\em et al.}, ``{\em {Search for a
  time-varying electron antineutrino signal at Daya Bay}},''
  \href{http://dx.doi.org/10.1103/PhysRevD.98.092013}{Phys. Rev. D {\normalfont
  \bfseries 98} (2018) no.~9, 092013},
  \href{http://arxiv.org/abs/1809.04660}{{\normalfont \ttfamily
  arXiv:1809.04660}}.

\bibitem{Borexino:2022khe}
{\normalfont \bfseries Borexino}, S.~Appel {\em et al.}, ``{\em {Independent
  determination of the Earth's orbital parameters with solar neutrinos in
  Borexino}},'' \href{http://arxiv.org/abs/2204.07029}{{\normalfont \ttfamily
  arXiv:2204.07029}}.

\bibitem{Super-Kamiokande:2003snd}
{\normalfont \bfseries Super-Kamiokande}, J.~Yoo {\em et al.}, ``{\em {A Search
  for periodic modulations of the solar neutrino flux in Super-Kamiokande
  I}},'' \href{http://dx.doi.org/10.1103/PhysRevD.68.092002}{Phys. Rev. D
  {\normalfont \bfseries 68} (2003)  092002},
  \href{http://arxiv.org/abs/hep-ex/0307070}{{\normalfont \ttfamily
  arXiv:hep-ex/0307070}}.

\bibitem{SNO:2005ftm}
{\normalfont \bfseries SNO}, B.~Aharmim {\em et al.}, ``{\em {A Search for
  periodicities in the B-8 solar neutrino flux measured by the Sudbury neutrino
  observatory}},'' \href{http://dx.doi.org/10.1103/PhysRevD.72.052010}{Phys.
  Rev. D {\normalfont \bfseries 72} (2005)  052010},
  \href{http://arxiv.org/abs/hep-ex/0507079}{{\normalfont \ttfamily
  arXiv:hep-ex/0507079}}.

\bibitem{SNO:2009ktr}
{\normalfont \bfseries SNO}, B.~Aharmim {\em et al.}, ``{\em {Searches for High
  Frequency Variations in the $^8$B Solar Neutrino Flux at the Sudbury Neutrino
  Observatory}},''
  \href{http://dx.doi.org/10.1088/0004-637X/710/1/540}{Astrophys. J.
  {\normalfont \bfseries 710} (2010)  540--548},
  \href{http://arxiv.org/abs/0910.2433}{{\normalfont \ttfamily
  arXiv:0910.2433}}.

\bibitem{Giunti:2021kab}
C.~Giunti, Y.~F. Li, C.~A. Ternes, and Z.~Xin, ``{\em {Reactor antineutrino
  anomaly in light of recent flux model refinements}},''
  \href{http://dx.doi.org/10.1016/j.physletb.2022.137054}{Phys. Lett. B
  {\normalfont \bfseries 829} (2022)  137054},
  \href{http://arxiv.org/abs/2110.06820}{{\normalfont \ttfamily
  arXiv:2110.06820}}.

\bibitem{Berryman:2021yan}
J.~M. Berryman, P.~Coloma, P.~Huber, T.~Schwetz, and A.~Zhou, ``{\em
  {Statistical significance of the sterile-neutrino hypothesis in the context
  of reactor and gallium data}},''
  \href{http://dx.doi.org/10.1007/JHEP02(2022)055}{JHEP {\normalfont \bfseries
  02} (2022)  055}, \href{http://arxiv.org/abs/2111.12530}{{\normalfont
  \ttfamily arXiv:2111.12530}}.

\bibitem{Goldhagen:2021kxe}
K.~Goldhagen, M.~Maltoni, S.~E. Reichard, and T.~Schwetz, ``{\em {Testing
  sterile neutrino mixing with present and future solar neutrino data}},''
  \href{http://dx.doi.org/10.1140/epjc/s10052-022-10052-2}{Eur. Phys. J. C
  {\normalfont \bfseries 82} (2022) no.~2, 116},
  \href{http://arxiv.org/abs/2109.14898}{{\normalfont \ttfamily
  arXiv:2109.14898}}.

\bibitem{Acero:2022wqg}
M.~A. Acero {\em et al.}, ``{\em {White Paper on Light Sterile Neutrino
  Searches and Related Phenomenology}},''
  \href{http://arxiv.org/abs/2203.07323}{{\normalfont \ttfamily
  arXiv:2203.07323}}.

\bibitem{Hagstotz:2020ukm}
S.~Hagstotz, P.~F. de~Salas, S.~Gariazzo, M.~Gerbino, M.~Lattanzi, S.~Vagnozzi,
  K.~Freese, and S.~Pastor, ``{\em {Bounds on light sterile neutrino mass and
  mixing from cosmology and laboratory searches}},''
  \href{http://dx.doi.org/10.1103/PhysRevD.104.123524}{Phys. Rev. D
  {\normalfont \bfseries 104} (2021) no.~12, 123524},
  \href{http://arxiv.org/abs/2003.02289}{{\normalfont \ttfamily
  arXiv:2003.02289}}.

\end{thebibliography}\endgroup

\end{document}